\documentclass[aps,prd,12pt,notitlepage,nofootinbib,tightenlines]{revtex4-1}
\usepackage{amsmath}
\usepackage{bm}
\usepackage{times}
\usepackage{braket}
\usepackage{color}
\usepackage{epsfig}
\usepackage{slashed}
\usepackage{hyperref}

\newcommand{\beq}{\begin{eqnarray}}
\newcommand{\eeq}{\end{eqnarray}}

\newcommand{\non}{\nonumber\\ }

\newcommand{\rmt}{ {\rm T}}
\newcommand{\kt}{ k_{\rm T} }
\newcommand{\psl}{ P \hspace{-2.5truemm}/ }

\newcommand{\etar}{ \eta^\prime }
\newcommand{\etap}{ \eta^{(\prime)} }

\newcommand{\epsl}{ \epsilon \hspace{-2.0truemm}/ }

\def \csb{Chin Sci  Bull }

\def \epja{ Eur Phys J A }
\def \epjc{ Eur Phys J C }
\def \jpg{  J Phys G }
\def \mpla{  Mod Phys Lett A }
\def \npb{  Nucl Phys B }
\def \plb{  Phys Lett B }
\def \pr{  Phys Rep }
\def \rmp{ Rev Mod Phys }
\def \prd{  Phys Rev D }
\def \prl{  Phys Rev Lett  }
\def \zpc{  Z Phys C  }
\def \jhep{ J High Energy Phys  }
\def \ijmpa{ Int J Mod Phys A }

\definecolor{Red}{rgb}{1.,0.,0.}

\definecolor{Blue}{rgb}{0.,0.,1.}

\definecolor{nicered}{rgb}{0.7,0.1,0.1}
\definecolor{nicegreen}{rgb}{0.1,0.5,0.1}

\hypersetup{colorlinks,citecolor=nicegreen,linkcolor=nicered}
\begin{document}
%%%%%%%%%%%%%%%%%%%%%%%%%%%%%%%%%%%%%%%%%%%%%%%%
\title{The semileptonic decays of $B/B_s$ meson in the perturbative QCD approach: A short review}
\author{Zhen-Jun Xiao$^{1,2}$\footnote{Email address: xiaozhenjun@njnu.edu.cn},
Ying-Ying Fan$^{1}$, Wen-Fei Wang$^{3}$ and Shan Cheng$^{1}$ }
\affiliation{1. Department of Physics and Institute of Theoretical Physics,
Nanjing Normal University, Nanjing, Jiangsu 210023, P.R. China,}
\affiliation{2. Jiangsu Key Laboratory for Numerical Simulation of Large Scale Complex Systems,
Nanjing Normal University, Nanjing 210023, People's Republic of China,}
\affiliation{3. Institute of High Energy Physics and Theoretical Physics Center
for Science Facilities, CAS, P.O.Box 918(4), Beijing 100049, P.R. China }
\date{\today}

%---------------------------------------------------------------------------------------------------------------%
\begin{abstract}
\noindent{\bf \large \hspace{-0.8cm} Abstract} \ \ \
In this short review, we present the current status about the theoretical and experimental studies
for some important semileptonic decays of $B/B_s$ mesons. We firstly gave a brief introduction for the
experimental measurements for $B/B_s \to P (l^+l^-, l^-\bar{\nu}_l, \nu \bar{\nu})$ decays, the
BaBar's $R(D)$ and $R(D^*)$ anomaly, the $P_5^\prime$ deviation for $B^0 \to K^{*0} \mu^+ \mu^-$ decay.
We then made a careful discussion about the evaluations for the relevant form factors
in the light-cone QCD sum rule (LCSRs),  the heavy quark effective theory,
and the perturbative QCD factorization approach.
By using the form factors calculated in the perturbative (pQCD) approach, we then calculate
and show the pQCD predictions for the decay rates of many semileptonic decays of $B/B_s$ mesons.
We also made careful phenomenological analysis for these pQCD predictions and found, in general,
the following points:
(a) For all the considered $B/B_s$ semileptonic decays, the next-to-leading order (NLO)
pQCD predictions for their decay rates agree well with the data
and those from other different theoretical methods;
(b) For $R(D)$ and $R(D^*)$, the pQCD predictions agree very well with the data,
the BaBar's anomaly of $R(D^{(*)})$ are therefore explained successfully
in the standard model by employing the pQCD approach;
and (c) We defined several new ratios $R_D^{l,\tau}$ and $R_{D_s}^{l,\tau}$, they may be more sensitive
to the QCD dynamics which controls the $B/B_s  \to (D^{(*)},D_s^{(*)} )$ transitions than the
old ratios, we therefore strongly suggest LHCb and the forthcoming Super-B experiments to measure these new ratios.
\end{abstract}

%%\pacs{13.20.He, 12.38.Bx, 14.40.Nd}

\maketitle
{\bf Key Words} $B/B_s$ meson semileptonic decays; The pQCD factorization approach;
           Form factors; Branching ratios; LHCb experiments

%---------------------------------------------------------------------------------------------------------------%

\section*{1. Introduction}\label{sec:1}

As is well-known, the semileptonic (SL) decays of $B$ and $B_s$ meson are
very important processes in testing the standard model (SM) and in searching
for the signal and/or evidence of the new physics (NP) beyond the standard model:
such as the extractions of the Cabbibo-Kobayashi-Maskawa matrix elements $|V_{ub}|$ and $|V_{cb}|$, and the determinations of the form factors
$F_{0,+,\rmt}(q^2)$ for the $B/B_s$ transitions to the pion, kaon or other light mesons
\cite{Fajfer13a,Ligeti13a,Gershon13a,Straub13a}.
Since the Spring of 2012, the BaBar's anomaly about the ratio ${\cal R}(D^{(*)})$
\cite{prl109-101802,prd8807}
invoked intensive studies for $B \to D^{(*)} l^-\bar{\nu}_l$ decays in the
framework of the
SM and various new physics (NP) models, for example,
in Refs.~\cite{prl109-071802,prd85-114502,prl109-161801,
prd86-054014,jhep2013-01054,prd86-034027,prd86-114037,mpla27-1250183,Fajfer-1301}.
The $B\to K^* \mu^+\mu^-$ anomaly observed by LHCb experiments
\cite{lhcb1308a,lhcb1308b,Serra13}
also stimulate many interesting studies \cite{alb2013,Jager1305,gmvprd88,as1308,ggh1308}.
We here will present a short review about the experimental
measurements and the theoretical studies
for the following $B/B_s$ semileptonic decay modes:
\beq
B/B_s &\to & P (l^+l^-,l\nu,\nu\bar{\nu} ), \quad (D^{(*)},D_s^{(*)})
l^-\bar{\nu}_l, \label{eq:bbs02}\\
B     &\to & K^* \mu^+\mu^-, \label{eq:b2k03}
\eeq
where $l=(e,\mu,\tau)$ are leptons and $P=(K, \pi, \eta,\etar)$ are light
pseudoscalar mesons.
For those considered $B/B_s \to P l\bar{\nu}$ decays, as illustrated in Fig.1,
the "Tree" Feynman diagrams provide the dominant leading order (LO) contribution.
For those $B/B_s \to P l^+ l^-$ and $P\nu \bar{\nu}$ decays, however,
the dominant LO standard model contributions come from those electroweak penguin
diagrams and $W^+W^-$ box diagrams.
For $B/B_s \to D_{(s)}^{(*)} l^-\bar{\nu}_l$ decays, the $b \to c l^- \bar{\nu}_l$
transition at the quark level provide the dominant contribution.

As for the relevant experimental measurements, some considered decays
of $B \to P( l^+l^-, l\nu, \nu \bar{\nu})$
have been measured by the Belle, BaBar, CLEO and/or  LHCb experiments
\cite{babar-83-032007,babar11,cleo-99-041802,belle-648-139,babar-86a,babar-86b,lhcb-2012a}.
The LHCb and the forthcoming Super-B experiments \cite{D-2012,Buras-2012} will
measure the $B_s \to P ( l^+l^-, l\nu, \nu \bar{\nu})$ decays in the near future.

AS for the $B \to D^{(*)} l\bar{\nu}_{\rm l}$ decays, they have been measured by both
BaBar and Belle collaboration \cite{babar2008,belle2007,belle2010}.
Very recently, the BaBar collaboration reported their measurements for
the ratios $R(D^{(*)})$ of the corresponding  branching
ratios \cite{prl109-101802,prd8807}:
\beq
\label{eq:exp02}
{\cal R}(D)& \equiv & \frac{{\cal B}(B \to D \tau^- \bar{\nu}_\tau)}{{\cal
B}(B \to D l^- \bar{\nu}_l)} = 0.440 \pm 0.072,\label{eq:rd01}\\
{\cal R}(D^*) & \equiv & \frac{{\cal B}(B \to D^* \tau^- \bar{\nu}_\tau)}{{\cal B}(B
\to D^* l^- \bar{\nu}_l)} = 0.332 \pm 0.030. \label{eq:rds01}
\eeq
These BaBar results are surprisingly larger than the SM predictions as
given, for example, in Ref.~\cite{prd85-094025}:
\beq
{\cal R}(D)^{\rm SM} &=&0.296\pm 0.016,\quad
{\cal R}(D^*)^{\rm SM} = 0.252\pm 0.003,
\label{eq:sm01}
\eeq
The combined BaBar results disagree with the SM predictions
by $3.4\sigma$ \cite{prl109-101802,bozek-2013}.
The type-II two-Higgs-doublet model (2HDM) with a charged-Higgs boson is excluded at
$99.8\%$ confidence level for any value of $\tan\beta/m_H$ \cite{prd8807}.

For the $B\to K^* \mu^+\mu^-$ decays, recent LHCb measurements show a good agreement
with the SM predictions for most physical observables \cite{lhcb1308a,lhcb1308b,Serra13}.
Some deviations of the angular observables from the SM have been
observed yet \cite{alb2013,Jager1305,gmvprd88,as1308,ggh1308}.
With $3.7\sigma$ the most significant discrepancy arises in the variable
$P_5^\prime$ \cite{dg1301}. Further LHCb studies based on more luminosity
will be necessary to clarify whether the observed deviations are a real sign
of NP or simply statistical artifacts \cite{Jager1305,gmvprd88}.

%%====================================================

On the theory side, we know that the central issues for the considered Ssemileptonic
$B/B_s$ decays are the estimations of the values and shapes of the
relevant form  factors for $B/B_s \to (P, D_{(s)}^{(*)})$ transitions.
The traditional methods or approaches to calculate the relevant transition form factors
are the light cone QCD sum rules (LCSR)
\cite{pball-98,pball-98b,kr00,huang01,wang03,pball-05,jhep04-014,zuo06,wu08,
wu09,huang09,prd83-094031,fu2013},
the heavy quark effective theory (HQET) \cite{Fajfer13a,Fajfer-1301,HQET1,HQET2,
HQET3,Grozin-2004}
and the lattice QCD (LQCD) techniques \cite{HPQCD-2006,Le2011,BL2013}.
In the pQCD approach, however,
one can make direct perturbative calculation for the form factors for
$B \to (\pi, K, etc)$ transitions
\cite{li-65,yang-npb642,yang-epjc23-28,lu-79,huang-71,wu07}.
Since the hadronic form factors always involve the non-perturbative QCD dynamics
\cite{krw98}, the QCD factorization approach \cite{npb592-3} based on the collinear
factorization can not be applied to compute the heavy-to-light form
factors directly, but take the soft form factors as input.

In the pQCD factorization approach \cite{li-95,li-96,li-97}, in fact,
one can write the form factors conceptually as a convolution of a hard kernel
with the distribution amplitudes of those mesons
involved in the decays.
Since the longitudinal momentum $k_{\rm L}$ approaches zero in the end point
region, the parton transverse momenta $k_{\rm T}$ here become non-negligible.
The resummation of the large double logarithmic term $\alpha_s \ln^2(k_{\rm T})$, or the large logarithms $\alpha_s \ln^2(x)$ can lead to the famous
Sudakov form factors \cite{Botts89,Catani89,Sterman92,Huang91,li-02,Cao95}.
Such Sudakov factors can strongly suppress the endpoint singularity,
which in turn can help one to make the perturbative calculation reliably.

In Refs.~\cite{li-65,lu-79}, for instance, Li et al. calculated the
form factors for $B \to (\rho, \pi)$ and $B \to S$ transitions
at the full leading order by using the pQCD approach
and found that their pQCD predictions for the corresponding form factors
are consistent well with those obtained by employing the light-cone sum rules or
other different theoretical methods \cite{pball-98,pball-98b,kr00,pball-05,jhep04-014,prd83-094031,cheng06,wal08,han2013}.
In Ref.~\cite{li-85074004}, very recently, Li, Shen and Wang calculated the
NLO twist-2 corrections to the $B \to \pi$ transition form factors at leading twist
(i.e. LO twist-2 contribution and LO twist-3 contribution ) in
the $k_{\rm T}$ factorization theorem. They found that
the NLO twist-2 contributions can amount up to $30\%$ of the value of the
form factors at the large recoil region of the pion. The calculation for the
NLO twist-3 contributions to the form factors of $B \to \pi$ transition
will be completed very soon \cite{cheng14a}.

%%%%%%%%%%%%%%%%%%%%%%%%%%%%%%%%%%%%%%%%%%%%%%%%
\begin{figure}[tbp]
\vspace{-6.5cm}
\centerline{\epsfxsize=14cm \epsffile{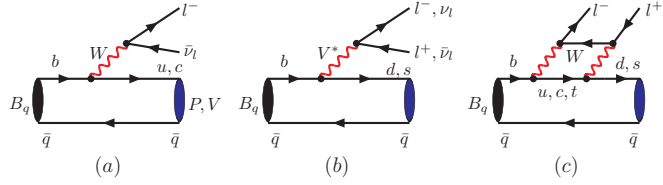} }
\vspace{-11cm}
\caption{ The typical Feynman diagrams for the SL decays
$B/B_s \to (P,V)(l^+l^-, l\bar{\nu}, \nu\bar{\nu})$ with $P=(\pi,K,\eta,\eta^\prime,D,D_s)$,
$V=(K^*, D^*,D_s^*)$ and $V^*=(\gamma^*,Z^*)$.}
\label{fig:fig1}
\end{figure}

%%%%%%%%%%%%%%%%%%%%%%%%%%%%%%%%%%%%%%%%%%%%%%%%

%---------------------------------------------------------------------------------------------------------------%
\section*{ 2. The form factors in the pQCD factorization approach} \label{sec:2}

As mentioned in last section, the central issue of the $B/B_s$ SL decays considered in
this paper are the evaluation of the relevant form factors of the $B \to (P, V)$ transitions,
in which $P=(\pi, K, \eta, \eta^\prime, D, D_s)$ are
pseudo-scalars and $V=(K^*, D^*,D_s^*)$ are the vector mesons.
In this section, we take $B \to \pi$ transitions as an example,
to show how to calculate the form factors in the pQCD factorization
approach. For more details for other cases, one can see the original
papers, for example, in Refs.~\cite{li-95,li-96,li-97,li-85074004,Xiao-SL1,Xiao-SL2,
Xiao-SL3,Xiao-SL4,Xiao-08a,Xiao-13a,Xiao-12a,keum2001,keum2001b,lu2001,xiao2006}.

\subsection*{2.1 The form factors $F_{0,+}(0)$ for $B\to \pi$ transition: One example}

In this paper, for the sake of simplicity,
we generally use $B$ to denote both $B$ and $B_s$ meson
and $P$ for the pseudo-scalar mesons, such as the pion, kaon and $\etap$, where
it is appropriate.
In the rest frame of B meson, we define the $B$ meson momentum $p_1$
and the final meson $P$ momentum $p_2$ in the light-cone coordinates:
$p_1=m_B(1,1,0_{\rm T})/\sqrt{2}$, $p_2=m_B\eta(0,1,0_{\rm T})/\sqrt{2}$,
where the parameter $\eta=1-q^2/m_B^2$ is the energy fraction of
the final state meson, and $q=p_1-p_2$ in the momentum carried by the final state
leptons. The momenta $k_1$ and $k_2$ are parameterized as those in
Ref.~\cite{Xiao-SL1}.

For the final state $\pi$ meson, we  adopt the distribution
amplitudes $\phi_\pi^A(x)$ (the twist-2 part ) and
$\phi_\pi^{P,T}(x)$ (the twist-3 part) as defined in
Refs.~\cite{pball-05,pball-06,pball-pi}:
\beq
\phi_\pi^A(x) &=& \frac{3 f_\pi}{\sqrt{6} }\, x(1-x) \left [ 1 + a_1^\pi
C_1^{3/2}(t) + a_2^\pi C_2^{3/2}(t)+a_4^\pi C_4^{3/2}(t)\right] \;,
\label{eq:phipik-a}\\
\phi^P_\pi(x) &=& \frac{f_\pi}{2\sqrt{6}}\, \left [ 1 +\left(30\eta_3
-\frac{5}{2}\rho_\pi^2\right) C_2^{1/2}(t)-3\left[
\eta_3\omega_3 + \frac{9}{20}\rho_i^2(1+6a_2^\pi ) \right ] C_4^{1/2}(t)
\right ]\;, \label{eq:phipik-p} \\
\phi^T_\pi(x) &=& \frac{f_\pi}{2\sqrt{6}}\, x(1-x) \left [ 1
+ \left(5\eta_3 -\frac{1}{2}\eta_3\omega_3 -
\frac{7}{20} \rho_\pi^2 - \frac{3}{5}\rho_\pi^2 a_2^\pi \right)C_2^{3/2}(t) \right ]\;,
\label{eq:phipik-t}
\eeq
where $t=2x-1$, $\rho_{\pi}=m_{\pi}/m_0^{\pi}$ is the mass ratios
with $m_0^\pi=1.4\pm 0.1$ GeV
is  the chiral mass of pion, $a_i^{\pi}$  are the Gegenbauer moments,
while $C_n^{\nu}(t)$ are the Gegenbauer polynomials \cite{Xiao-SL1}.
The values of $a_i^{\pi,K}$
can be found in Eq.~(13) of Ref.~\cite{Xiao-SL1}.
It is worth of mentioning that a new progress
about pion form factor in the $\pi \gamma^* \to \gamma$ scattering
has been made in Ref.~\cite{jhep1401-004} very recently, where the authors made
a joint resummation for the pion wave function and the pion transition form factor
and proved that the $\kt$ factorization is scheme independent.

For the wave functions of the B and $B_s$ meson, there are a lot of studies for
their structure and shapes, form example,  in the framework of the heavy quark
limit \cite{qiao01,huang05,huang06}.
In Ref.~\cite{jhep1302-008}, the authors studied the rapidity resummation
improved $B$ meson wave function and found that the resummation effect
keeps the normalization of the B meson wave functions
and strengths their convergent behavior at small spectator momentum.
For more details about the wave functions of $B/B_s$ meson, one can see a new review paper \cite{wh2013}
and references therein.
We here still use the $B/B_s$ wave functions as defined in
Refs.~\cite{li-65,yang-npb642,yang-epjc23-28,lu-79}.
For the distribution amplitudes (DA's) of $B/B_s$ meson, we adopt the same form
as being used in Refs.~\cite{Xiao-SL1,Xiao-08a,Xiao-13a,Xiao-12a,keum2001,keum2001b}:
\beq
\phi_{B}(x,b)&=& N_Bx^2(1-x)^2 \exp\left[-\frac{1}{2}\left(\frac{x m_B}{\omega_B}\right)^2
-\frac{\omega_B^2 b^2}{2}\right],\label{eq:b01}
\eeq
where the normalization factors $N_B$ ($N_{B_s}$) are related to the decay
constants $f_B$ ($f_{B_s}$ ) through the normalization relation
$\int_0^1 dx \phi_{B_{(s)}}(x, b=0) =f_{B_{(s)}}/(2\sqrt{6})$.
The shape parameter $\omega_B=0.40\pm 0.04$ GeV
and $\omega_{B_s} = 0.50 \pm 0.05$~GeV were estimated by using the
rich experimental measurements and setting
$f_{B}= 0.21$~GeV and $f_{B_s} = 0.23$~GeV.

%%%%%%%%%%%%%%%%%%%%%%%%%%%%%%%%%%%%%%%%%%%%%%%%
\begin{figure}[tbp]
\vspace{-4cm}
\centerline{\epsfxsize=15cm \hspace{2cm} \epsffile{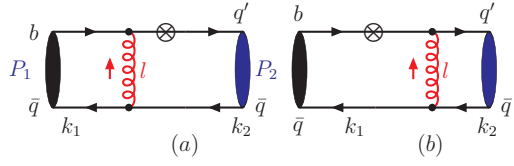} }
\vspace{-14.5cm}
\caption{ The Feynman diagrams responsible for the extraction of the
LO twist-2 and LO twist-3 contributions to the form factor of $B \to \pi$ transition.
The symbol $\otimes$ refers to the weak vertex.}
\label{fig:fig2}
\end{figure}

%%%%%%%%%%%%%%%%%%%%%%%%%%%%%%%%%%%%%%%%%%%%%%%%

The form factors $F_{0,+}(q^2)$ and $F_{\rm T}(q^2)$ for $B\to P$
transitions with $P=\pi$ or $K$ are defined
in the usual way as in  Ref.~\cite{pball-05,pball-06,pball-pi}.
In order to cancel the poles at $q^2=0$, $F_+(0)=F_0(0)$ must be satisfied.
For the sake of convenience, one usually define $F_{0,+}(q^2)$ as a summation of the
auxiliary form factors $f_1(q^2)$ and $f_2(q^2)$:
\beq
F_+(q^2)&=&\frac12[f_1(q^2)+f_2(q^2)],  \label{eq:ffp01} \\
F_0(q^2)&=&\frac12 f_1(q^2) \left [1+\frac{q^2}{m_B^2-m_P^2} \right ]
+\frac12 f_2(q^2)\left [1-\frac{q^2}{m_B^2-m_P^2} \right ].
\label{eq:ffp02}
\eeq

In the pQCD approach, one can calculate perturbatively the LO twist-2 and LO twist-3
contributions to the form factors, through the analytical calculations for the two
factorizable emission Feynman diagrams as shown in Fig.~2.
By taking the Sudakov form factors and the threshold resummation
effects into account, we calculated and found the form factors
$f_{1,2}(q^2)$ and $F_{\rm T}$ for $B \to P$ transitions,
as given for example in Eqs.~(19-21) of Ref.~\cite{Xiao-SL1}.

In the pQCD approach at LO level, the form factors $F_{0,+}(q^2)$ and $F_{\rm T}(q^2)$
as defined in Eqs.~(\ref{eq:ffp01},\ref{eq:ffp02})
include the LO twist-2 and LO twist-3 contributions only. They, of course, are the
dominant part of the form factors in consideration \cite{li-85074004}.
In Ref.~\cite{li-85074004}, the authors compared the relative strength of the LO twist-2 part and LO twist-3 part, and
then calculated the NLO twist-2 contribution to the form factors
of $B \to \pi$ transition, found the corresponding NLO factor $F_{\rm twist-2}=F(x_1,x_2,\eta,\mu_f,\mu,\zeta_1)$. The
explicit expression of the NLO factor  $F(x_1,x_2,\eta,\mu_f,\mu,\zeta_1)$ can be found easily in Refs.~\cite{li-85074004,Xiao-SL1}.
At the NLO level, consequently, the NLO hard kernel $H$ can be written as the form of ~\cite{li-85074004}
\beq
H=H^{(0)}(\alpha_s)+ H^{(1)}(\alpha_s^2)
= \left [ 1+F_{\rm twist-2}(x_1,x_2,\mu,\mu_f,\eta,\zeta_1) \right ] H^{(0)}(\alpha_s),
\eeq
where the hard kernel $H^{(0)}(\alpha_s)$ contains the LO twist-2 and LO twist-3 contributions.
The NLO twist-3 contribution  $F_{\rm twist-3}$ is another part of the NLO contribution to the form
factors in the framework of the pQCD factorization approach,
which is still absent now but in the
process of analytical calculation \cite{cheng14a}. Based on the $SU(3)_F$ flavor
symmetry, we can find the similar expressions for the form factors of other final
state pseudoscalar mesons, such as $K, \eta$ and $\etar$ meson\cite{Xiao-SL1,Xiao-SL2}.

\subsection*{2.2 The form factors for $B\to D_{(s)}, D^*_{(s)}$ transitions}

For the pseudoscalar $D$ meson and the vector $D^*$ meson, their wave functions can be chosen
as \cite{prd78-014018,Xiao-SL3}
\beq
\Phi_{\rm D}(p,x)&=&\frac{i}{\sqrt{6}}\gamma_5 (\psl_{\rm D}+ m_{\rm D} )\phi_{\rm D}(x), \label{eq:wfd} \\
\Phi_{\rm D^*}(p,x) &=& \frac{-i}{\sqrt{6}} \left [  \epsl_{\rm L}(\psl_{\rm D^*} +m_{\rm D^*})\phi^L_{\rm D^*}(x)
 + \epsl_{\rm T}(\psl_{\rm D^*} + m_{\rm D^*})\phi^T_{\rm D^*}(x)\right ]
\label{eq:wfdstar}.
\eeq
For the distribution amplitudes of $D^{(*)}$ meson, we adopt the one as defined in Ref.~\cite{prd78-014018}
\beq
\phi_{\rm D^{(*)} }(x)=\frac{ f_{\rm D^{(*)}}}{2\sqrt{6}} 6x(1-x) \left[ 1+C_{D^{(*)} }(1-2x)\right]
\cdot \exp \left[-\frac{\omega^{\rm 2} b^{\rm 2} }{2}\right].
 \label{eq:phid}
\eeq
From the heavy quark limit,  we here assume that $f^{\rm L}_{\rm D^*}=f^{\rm T}_{\rm D^*}
=f_{\rm D^*}$,~$\phi^{\rm L}_{\rm D^*}=\phi^{\rm T}_{\rm D^*}=\phi_{\rm D^*}$, and set $C_{\rm D}=C_{\rm D^*}=0.5,~
\omega=0.1$ GeV as Ref.~\cite{prd78-014018,Xiao-SL3}.

For $B\to D$ transition,  the form factors $F_{\rm 0,+}(q^{\rm 2})$
can be written in terms of $f_{\rm 1,2}(q^{\rm 2})$ as in Eq.~(\ref{eq:ffp01},\ref{eq:ffp02}). The explicit expressions
of $f_{1,2}(q^2)$ for $B\to D$ transition can be found easily in Eqs.~(14-18) of Ref.~\cite{Xiao-SL3}.
For $B \to D^*$ transitions, the relevant form factors are $V(q^{\rm 2})$ and $A_{\rm 0,1,2}(q^{\rm 2})$ \cite{prd65-014007},
and have been given explicitly in Eqs.~(20-23) of Ref.~\cite{Xiao-SL3}.
For $B_s\to (D_s, D_s^*)$ transitions, the explicit expressions of the form factors
$F_{\rm 0,+}(q^2)$, $V(q^2)$ and $A_{\rm 0,1,2}(q^2)$ can be found directly in
Eqs.(17-19,24-27) of  Ref.~\cite{Xiao-SL4}.

\subsection*{2.3 The extrapolation of the form factors}

As mentioned in previous section, the central issue for the theoretical calculations
of the semileptonic $B/B_s$ decays are the evaluation of the values and the shape of
the relevant form factors $F_{\rm 0,+}(q^{\rm 2})$, $V(q^{\rm 2})$ and $A_{\rm 0,1,2}(q^{\rm 2})$.
For the $B/B_s \to P$ transition with $P=(\pi, K, etc)$ the light pseudoscalar mesons,
the two traditional  methods of evaluating the form factors are the LCSR in the low $q^{\rm 2}$ region
and the Lattice QCD for the high $q^{\rm 2}$ region of $q^{\rm 2}\approx q_{\rm max}^{\rm 2}$.
For the form factors of $B \to \pi, K$ transitions,
the relevant experiments also provide some help
to determine their value and the shape \cite{Bobeth-12,Gallo-08}.
The pQCD predictions for values of those form factors in low $q^2$ region
are consistent well with those from
LCSR \cite{Xiao-SL1,Xiao-SL2,prd78-014018,prd65-014007}.

For $B \to (D, D^*)$ transitions, the traditional methods to evaluate the form factors
are the HQET  \cite{Fajfer13a,Fajfer-1301,HQET1,HQET2,HQET3,Grozin-2004,BLT-12} in the low
$q^2$ region
and the LQCD techniques \cite{HPQCD-2006,Le2011,BL2013} in the high $q^{\rm 2}$ region.
In Refs.~\cite{li1995,prd67-054028,prd78-014018}, the authors examined the
applicability of the pQCD approach to
$B \to (D, D^*)$ transitions,
and have shown that the pQCD approach with the inclusion of the Sudakov
effects is  applicable to the $B \to D^{(*)} l\bar{\nu}_{\rm l}$ decays
in the lower $q^{\rm 2}$ region ( i.e. the $D$ or $D^*$ meson recoils fast).
Since the pQCD predictions for the relevant form factors
are reliable in the low $q^{\rm 2}$ region only, we will calculate explicitly
the values of the form factors $F_{\rm 0,+}(q^{\rm 2})$, $V(q^{\rm 2})$
and $A_{\rm 0,1,2}(q^{\rm 2})$ in the lower range of
$ m_{\rm l}^{\rm 2} \leq q^{\rm 2} \leq m_{\rm \tau}^{\rm 2}$
with $l=(e,\mu)$ by  using the expressions as given in previous subsection.

In the low $q^{\rm 2}$ region of $m_l^2 \leq q^{\rm 2} \leq m_{\rm \tau}^{\rm 2}$,
we firstly calculate the form factors $F_i(q^{\rm 2})$ for $B \to P,D^{(*)}_{(s)}$
transitions at some points by employing the pQCD approach respectively.
Secondly we make an extrapolation for the form factors $F_i(q^{\rm 2})$
from the low $q^{\rm 2}$ region to the high $q^{\rm 2}$ region.
In Refs.~\cite{Xiao-SL1,Xiao-SL2}, we use different parametrization
for $F_0(q^2)$ and $F_{+,T}(q^2)$ respectively.
For the form factor $F_0(q^2)$ of $B/B_s\to (\pi, K)$ transitions,
we us the classical pole model parametrization to make the extrapolation
\beq
 F_0(q^2)=\frac{F_0(0)}{1-a(q^2/m_B^2)+b(q^2/m_B^2)^2}, \label{eq:pole-1}
\eeq
where the parameter $a$ and $b$ will be determined by the fitting procedure
as described in Refs.~\cite{Xiao-SL1,Xiao-SL2}.

For $F_{\rm +,T}(q^2)$, we use the Ball/Zwicky(BZ) parametrization
to do the extrapolation \cite{pball-05,pball-98,pball-98b,kr00,param-bz2}
\beq
\label{pa-bz}
F_i(q^2) = \frac{F_i(0)}{1-q^2/m_{B_{(s)}^*}^2}
 + \frac{ F_i(0) \;r q^2/m_{B_{(s)}^*}^2}{\left(1-q^2/m_{B_{(s)}^*}^2\right)
\left(1-\alpha\,q^2/m_{B_{(s)}}^2\right)},
\eeq
where the shape parameters $\alpha$ and $r$ could be determined by the
fitting procedure the same as in Ref.~\cite{Xiao-SL1,Xiao-SL2}.

%%------------------------------------------------
\begin{table}[thb]
\begin{center}
\caption{The pQCD predictions for the form factors $F_{0,+}(0)$
and $F_{\rm T}(0)$
for $B/B_s \to (\pi,K)$ transitions, and $F_{0,+}(q^2)$, $V(q^2)$ and $A_{0,1,2}(q^2)$
for $B\to D^{(*)}$  transitions with $q^2=0, m_\tau^2$, respectively. \label{tab-f0}}
\vspace{0.5cm}
\begin{tabular}{l l l l}\hline\hline
Transitions    &$ F_0(0)_{\rm LO}$     &$F_+(0)_{\rm LO}$         &$F_{\rm T}(0)_{\rm LO}$  \\  \hline
$B\to\pi$\;    &$0.22^{+0.04}_{-0.03}$ &$0.22^{+0.04}_{-0.03}$\;  &$0.23^{+0.04}_{-0.04}$       \\  \hline
$B\to K$       &$0.27^{+0.05}_{-0.04}$ &$0.27^{+0.05}_{-0.04} $   &$0.30^{+0.05}_{-0.04}$   \\   \hline
$B_s\to K$\;   &$0.22\pm 0.04$         &$0.22 \pm 0.04$           &$0.25^{+0.05}_{-0.04}$   \\ \hline\hline
%%--------------------------------------------------------------------------
               & $F_0(0)_{\rm NLO}$      &$F_+(0)_{\rm NLO}$     &$F_{\rm T}(0)_{\rm NLO}$    \\  \hline
$B\to\pi$\;    &$0.26^{+0.05}_{-0.04}$\; &$0.26^{+0.05}_{-0.04}$ &$0.26^{+0.05}_{-0.04}$    \\  \hline
$B\to K$       &$0.31 \pm 0.05$          &$0.31\pm 0.05$         &$0.34^{+0.06}_{-0.05}$      \\   \hline
$B_s\to K$\;   &$0.26^{+0.05}_{-0.04}$   &$0.26^{+0.05}_{-0.04}$ &$0.28^{+0.06}_{-0.06}$      \\ \hline\hline
%%--------------------------------------------------------------------------
                   &$ F_0(0)$              &$F_+(0)$                 &$V(0)$                  \\  \hline
$B\to D^{(*)}$     &$0.52^{+0.12}_{-0.10}$ &$0.52^{+0.12}_{-0.10}$\; &$0.59^{+0.12}_{-0.11}$  \\  \hline
                   &$ F_0(m_\tau^2)$       &$F_+(m_\tau^2)$          &$V(m_\tau^2)$            \\  \hline
$B\to D^{(*)}$     &$0.64^{+0.14}_{-0.12}$ &$0.70^{+0.16}_{-0.14}$\; &$0.79^{+0.15}_{-0.14}$  \\  \hline \hline
                   &$A_0(0)$               &$A_1(0)$                 &$A_2(0)$                \\  \hline
$B\to D^{(*)}$     &$0.46^{+0.10}_{-0.08}$ &$0.48^{+0.10}_{-0.09}$   &$0.51^{+0.11}_{-0.09}$  \\  \hline
                   &$A_0(m_\tau^2)$        &$A_1(m_\tau^2)$          &$A_2(m_\tau^2)$         \\  \hline
$B\to D^{(*)}$     &$0.62^{+0.12}_{-0.11}$ &$0.58^{+0.11}_{-0.10}$   &$0.66^{+0.13}_{-0.12}$  \\  \hline\hline
%%--------------------------------------------------------------------------
\end{tabular}
\end{center}
\end{table}

%%====================================

In Table \ref{tab-f0}, we collect the LO and NLO pQCD predictions for the transition
form factors $F_{0,+}(0)$ and $F_{\rm T}(0)$ for the considered decay modes.
The total errors are obtained by adding the individual errors
in quadrature.  In this table, we also
show the LO pQCD predictions for the form factors $F_{0,+}(q^2)$, $V(q^2)$ and
$A_{0,1,2}(q^2)$ for $B \to D^{(*)}$ transitions with $q^2=0, m_\tau^2$, respectively.
For more details see Refs.~\cite{Xiao-SL1,Xiao-SL2,Xiao-SL3,Xiao-SL4}.
One can see from the theoretical predictions as listed in Table \ref{tab-f0} that
the pQCD predictions for the
form factors of $B\to (\pi, K, D)$ and $B_s \to K$ transitions at $q^2=0$
generally agree well with those from LCSRs \cite{jhep1009-089,pball-05}
within one standard deviation.

\begin{figure*}
\centering
\includegraphics[width=0.32\textwidth]{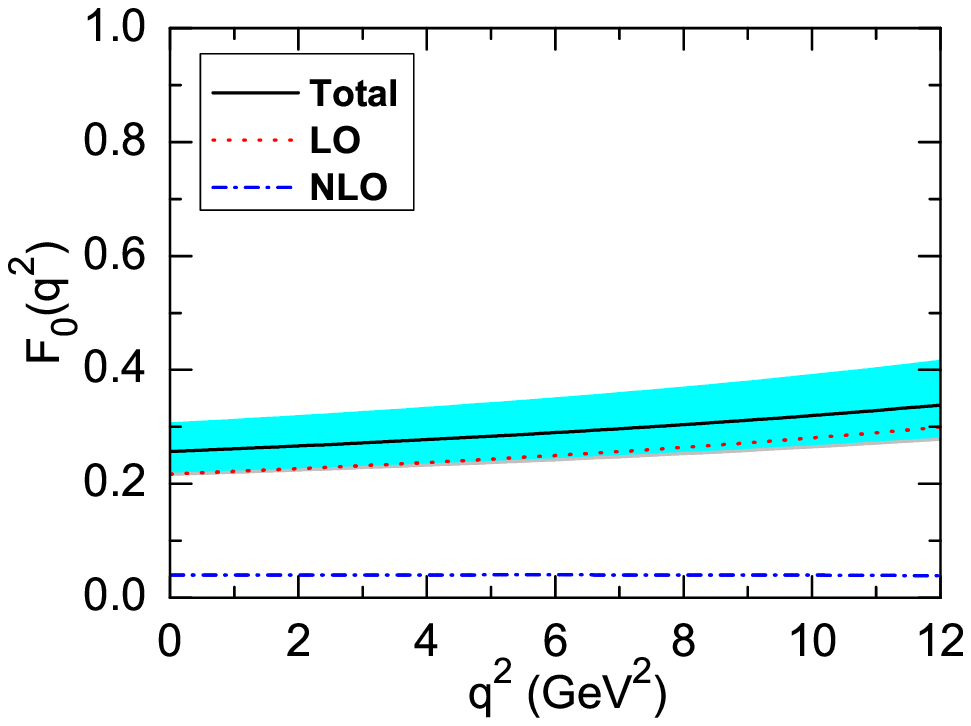}
\includegraphics[width=0.32\textwidth]{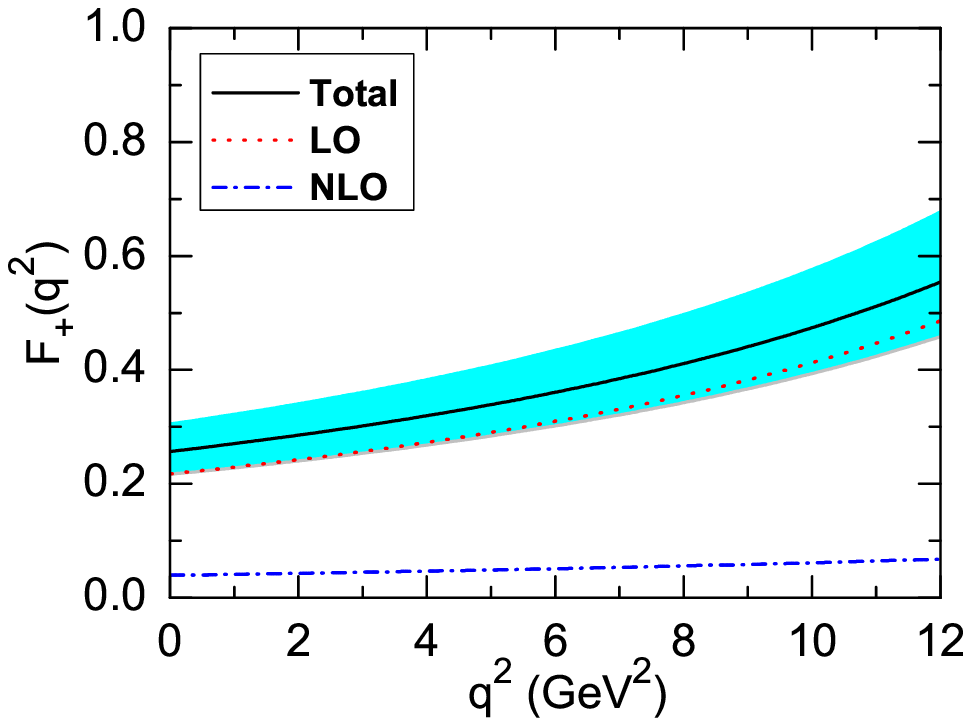}
\includegraphics[width=0.32\textwidth]{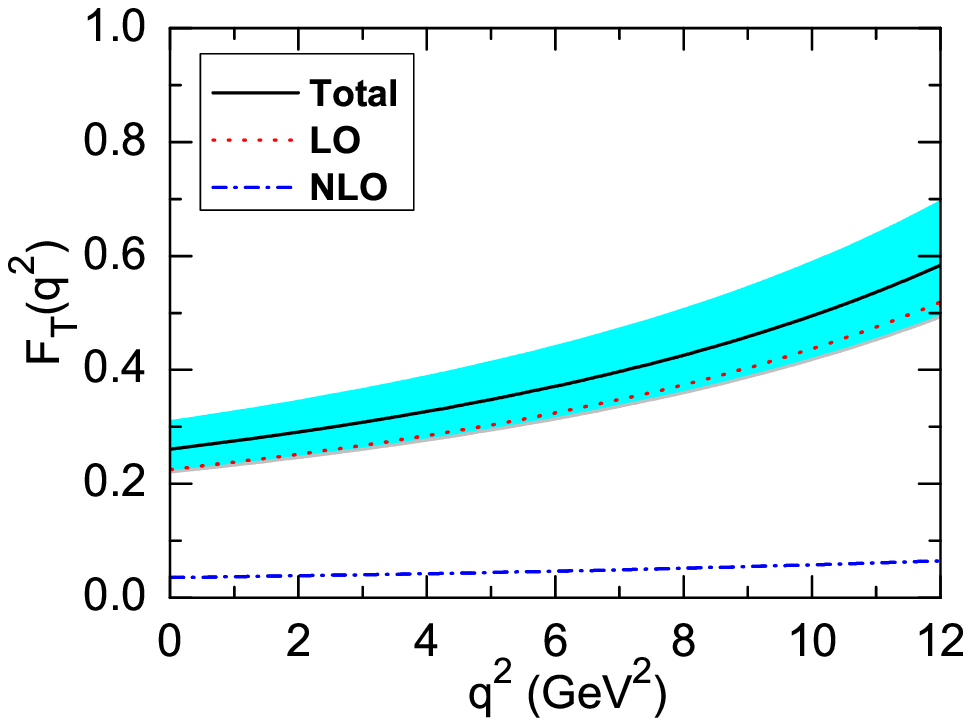}
\caption{The pQCD predictions for the form factors $F_{0,+,\rmt}(q^2)$
for $B\to \pi$ transition. The solid line denotes the total value
of the NLO results and the shaded band describes the total theoretical error.}
\label{fig:fig3}
\end{figure*}

In Fig.~\ref{fig:fig3}, as an example, we illustrate the $q^2$-dependence
of the pQCD predictions for the form factors
$F_{0,+,\rmt}(q^2)$ at the LO (dots lines) and the NLO (solid line)
for the $B \to \pi$ transition.
The shaded band in Fig.~3 illustrates the total error of the pQCD predictions
obtained by adding the different theoretical errors in quadrature.
For more details about the pQCD
predictions for the values and the $q^2$-dependence of the relevant  form
factors for other $B/B_s$ semileptonic decays
considered in this paper, one can see Refs.~\cite{Xiao-SL1,Xiao-SL2,Xiao-SL3,Xiao-SL4}.

%%-----------------------------------------------
%%----------------------------------------------

\subsection*{2.4 Form factors of $B \to D^{(*)}$ transitions in the HQET}\label{subsec:hqet}

The HQET is the traditional method for the evaluations of the form factors
for $B \to (D,D^*)$  transitions \cite{Fajfer13a,Fajfer-1301,HQET1,HQET2,HQET3}.
We here present the formulae for evaluating the form factors for
$B \to D^{(*)} l \bar{\nu}_l$ decays, quoted directly from
Refs.~\cite{prd85-094025,jhep2013-01054}.
\beq
F_+(q^2) &=&  \frac{m_B + m_D}{2\sqrt{ m_B m_D }} G_1(w),  \label{eq:ff11} \\
F_0(q^2) &=&   \frac{\sqrt{ m_B m_D }}{m_B + m_D} G_1(w)\; \Delta(w)\;
(1 + w)\frac{1+r}{1-r}, \label{eq:ff12}
\eeq
with the function $G_1(w)$ in the form of
\beq
G_1(w)=G_1(1)\left [ 1- 8 \rho_1^2\, z(w) + (51 \rho_1^2 -10)\, z(w)^2
 - (252 \rho_1^2 - 84)\, z(w)^3 \right],
\eeq
where $r=m_{D}/m_{B}$, $z(w)=(\sqrt{w+1}-\sqrt{2})/(\sqrt{w+1}+\sqrt{2})$,
the new kinematical variable $w$ is defined as
$w = v_B \cdot v_{D^{(*)}} = (m_{B}^2 + m_{D^{(*)}}^2
- q^2)/(2 m_{B} m_{D^{(*)}})$ with $q^2=(p_B-p_{D^{(*)}})^2$.
The scalar density $\Delta(w)$ is approximated by a constant value
$\Delta(w) = 0.46 \pm 0.02$ \cite{prd85-094025,jhep2013-01054}.
From Refs.~\cite{prd85-094025,jhep2013-01054}, we also find
\beq
G_1(1)|V_{cb}|&=& (42.64\pm 1.53)\times 10^{-3}, \non
\rho_1^2&=&1.186\pm 0.036\pm 0.041.
\eeq

For the $\bar B \to D^*$ transition, the form factors
$V(q^2)$ and $A_{0,1,2}(q^2)$ are related to the universal HQET
form factor $h_{A_1}(w)$ via~\cite{prd85-094025,CLN-97}
\beq
V(q^2)   &=& \frac{R_1(w)}{R_{D^*}}\, h_{A_1}(w) ,\quad
A_0(q^2) = \frac{R_0(w)}{R_{D^*}}\, h_{A_1}(w) \,, \non
A_1(q^2) &=& R_{D^*}\, \frac{w +1}{2}\, h_{A_1}(w) \,, \non
A_2(q^2) &=& \frac{R_2(w)}{R_{D^*}}\, h_{A_1}(w) \,,
\eeq
where $R_{D^*}=2\sqrt{m_B m_{D^*}}/(m_B+m_{D^*})$, while
$h_{A_1}(w)$ and ratios $R_{0,1,2}(w)$ are of the following
~\cite{prd85-094025,CLN-97}
\beq
h_{A_1}(w)&=& h_{A_1}(1)\,\left [1 - 8 \rho^2 z(w) + (53 \rho^2 - 15)\,z(w)^2
 -(231 \rho^2 - 91)\, z(w)^3 \right] \,, \non
R_0(w) &=& R_0(1) - 0.11(w-1) + 0.01(w-1)^2 \,, \non
R_1(w) &=& R_1(1) - 0.12(w-1) + 0.05(w-1)^2 \,, \non
R_2(w) &=& R_2(1) + 0.11(w-1) - 0.06(w-1)^2 \,.
\eeq
The parameters $\rho^2$, $R_1(1)$ and $R_2(1)$ are determined from
the well-measured $\bar B \to D^* \ell  \bar\nu$ decay distributions~\cite{hfag2012}
($\ell=e,\mu$),
\beq
\rho^2&=&1.207\pm 0.026, \quad R_1(1)=1.403\pm 0.033, \non
R_2(1)&=&0.854\pm 0.020, \quad R_3(1)=0.97\pm 0.10, \non
h_{A_1}(1)|V_{cb}|&=&(35.90\pm 0.45)\times 10^{-3}.
\eeq
While the parameter $R_0(1)$ can be derived from the equation
\beq
\frac{R_2(1) (1-r) + r\, [ R_0(1) (1 + r) - 2 ]}{(1-r)^2}
=R_3(1).
\eeq

%%==============================================================================

\section*{3.\ \   $B_{(s)} \to (\pi, K, \etap)(l^+l^-,l\nu,\nu\bar{\nu})$
decays:  Branching ratios} \label{sec:c3}

\subsection*{3.1 The formulae of differential decay widths}\label{sec:c3-1}

For the Semileptonic decays $B \to \pi l^- \bar{\nu}_l$ and $\bar{B}^0_s\to K^+ l^- \bar{\nu}_l$,
the quark level transitions are the $b\to ul^-\bar{\nu}_l$
with $l^-=(e^-,\nu^-, \tau^-)$, and the corresponding effective Hamiltonian
is of the form \cite{Buras96}
\beq
{\cal H}_{eff}(b\to ul\bar \nu_l)=\frac{G_F}{\sqrt{2}}V_{ub}
\cdot \bar{u} \gamma_{\mu}(1-\gamma_5)b \cdot \bar l\gamma^{\mu}(1-\gamma_5)\nu_l,
\eeq
where $G_F=1.166 37\times10^{-5} GeV^{-2}$ is the Fermi coupling constant,
$V_{ub}$ is one of the Cabbibo-Kobayashi-Maskawa quark mixing
matrix elements. The differential decay rates can be written as
\cite{lu-79,Col-2006}
\beq
\frac{d\Gamma(b\to ul\bar \nu_l)}{dq^2}&=&\frac{G_F^2|V_{ub}|^2}{192 \pi^3  m_B^3}
\left ( 1-\frac{m_{\rm l}^{\rm 2}}{q^{\rm 2}} \right)^{\rm 2}\frac{\lambda^{1/2}(q^{\rm 2})}{2q^{\rm 2}}
\non
&& \cdot \Bigl \{  3 m_{\rm l}^{\rm 2}\left (m_{\rm B}^{\rm 2}-m_{\rm P}^{\rm 2}\right )^{\rm 2}
 |F_{\rm 0}(q^{\rm 2})|^{\rm 2}
 + \left (m_{\rm l}^{\rm 2}+2q^{\rm 2} \right )\lambda(q^{\rm 2})|F_+(q^{\rm 2})|^{\rm 2} \Bigr \},
\label{eq:dg01}
\eeq
where $m_l$ is the mass of lepton, and
$\lambda(q^2) = (m_B^2+m_P^2-q^2)^2 - 4 m_B^2 m_P^2$ is the phase-space factor.

The effective Hamiltonian  for $b\to q l^-l^+ $ and $b \to d \nu \bar \nu$ transitions are
of the form
\beq
{\cal H}_{\mbox{eff}}&=& -\frac{G_F}{\sqrt{2}}V_{tb}V^*_{tq} \sum_{i=1}^{10}C_i(\mu)O_i(\mu),\\
{\cal H}_{b \to q\nu \bar \nu}&=& \frac{G_F}{\sqrt{2}} \frac{\alpha_{em}}{2 \pi
\sin^2(\theta_W)} V_{tb} V_{tq}^* \eta_X X(x_t)
\cdot \left[ {\bar q}\gamma^\mu (1-\gamma_5) b \right ] \left[ {\bar \nu} \gamma_\mu(1-\gamma_5) \nu \right],
\eeq
where $q=(d,s)$, $C_i(\mu)$ are the Wilson coefficients at the scalae $\mu$
and $O_i(\mu)$ are the local four-fermi operators \cite{Buras96,Xiao-SL1}.
The explicit expressions of the differential decay widths for these two kinds
of decays can be found easily in Refs.~\cite{Buras96,lu-79,prd81}.

\subsection*{3.2 pQCD predictions for $Br(B_{(s)} \to P (l^+l^-,l\nu,\nu\bar{\nu}))$ }\label{sec:c3-2}

Since the input parameters used in different papers may be a little different,
one can see individual papers for the choices of the input pararmeter about the
masses, decay constants, life-times and CKM elements \cite{pdg2012,hfag2012}.
By using the formula and the input parameters as given in previous sections,
we firstly calculate the branching ratios for the considered charged and neutral
current semileptonic decays.

After the numerical integration for $q^2$ over the range of $m_l^2 \leq q^2 \leq
(M_B-m_i)^2$, we obtain the pQCD predictions for the branching ratios of
all decay modes in consideration.
For $B/B_s \to (\pi, K)(l^+l^-,l\nu,\nu\bar{\nu})$ decays, all numerical results
are listed in Table \ref{tab-brn1}. The theoretical predictions obtained by
employing other methods also in the framework of the SM \cite{Bartsch09,choi10,wang08b}, as
well as the measured values currently available
\cite{babar-83-032007,babar11,cleo-99-041802,
belle-648-139,babar-86a,babar-86b,lhcb-2012a,lhcb-prd86a,pdg2012,hfag2012}, are all
included in this Table.
The total theoretical error of the pQCD predictions for the branching ratios are obtained by the
the combination in quadrature of the individual errors from $\omega_B$ or $\omega_{B_s}$,
$f_B$ or $f_{B_s}$, relevant Gegenbauer moments $a_i$ and the chiral mass $m_0^{\pi,K}$.

%%===============================================================
\begin{table}[thb]
\begin{center}
\caption{The pQCD predictions for the branching ratios of those considered
decay modes with $l=(e,\mu)$. The theoretical predictions
based on other methods \cite{Bartsch09,choi10,wang08b} and currently available data or the world averages
\cite{pdg2012,hfag2012,lhcb-prd86a}. The upper limits are at the $90\%$C.L.}
\label{tab-brn1}
\begin{tabular}{l| l|l  |l}  \hline\hline
 Decay modes       & pQCD& Others & Data \\  \hline
$Br(\bar{B}^0\to\pi^+ l^- \bar{\nu}_l)(10^{-4})$     & $1.42^{+0.52}_{-0.44} $ & & $ 1.44^{+0.05}_{-0.05}$  \\
$Br(\bar{B}^0\to\pi^+ \tau^-\bar{\nu}_\tau)(10^{-4})$& $0.90^{+0.33}_{-0.27}$ & &  \\
$Br(B^-\to\pi^0 l^- \bar{\nu}_l)(10^{-4})$     & $0.76^{+0.28}_{-0.24} $ & &$0.77^{+0.05}_{-0.05}$  \\
$Br(B^-\to\pi^0 \tau^-\bar{\nu}_\tau)(10^{-4})$& $0.49^{+0.18}_{-0.15} $ & &  \\ \hline\hline
$Br(\bar{B}^0\to\pi^0 l^+ l^-)(10^{-8}) $       & $0.91^{+0.34}_{-0.29}$ & & $< 12$  \\
$Br(\bar{B}^0\to\pi^0 \tau^+ \tau^-)(10^{-8})$  & $0.28^{+0.09}_{-0.09}$ &  &  \\
$Br(\bar{B}^0\to\pi^0 \nu  \bar{\nu})(10^{-8})$ & $7.30^{+2.72}_{-2.24}$ && $<22000$  \\ \hline
$Br(B^-\to\pi^- l^+ l^-)(10^{-8})$       & $1.95^{+0.73}_{-0.60}$ & $2.0^{+0.2}_{-0.2}$& $2.3\pm 0.6$  \\
$Br(B^-\to\pi^- \tau^+ \tau^-)(10^{-8})$  & $0.60^{+0.21}_{-0.17}$& &   \\
$Br(B^-\to\pi^- \nu  \bar{\nu})(10^{-8})$ & $15.7^{+5.8}_{-4.3}$ && $< 10000$  \\ \hline\hline
$Br(\bar{B}^0\to\bar{K}^0 l^+ l^-)(10^{-7})$        & $5.12^{+1.91}_{-1.57}$ && $4.7^{+0.6}_{-0.2}$  \\
$Br(\bar{B}^0\to\bar{K}^0 \tau^+ \tau^-)(10^{-7})$  & $1.20^{+0.42}_{-0.35}$ & &  \\
$Br(\bar{B}^0\to\bar{K}^0 \nu  \bar{\nu})(10^{-6})$ & $4.11^{+1.54}_{-1.26}$ && $ < 56 $  \\ \hline
$Br(B^-\to K^- l^+ l^-)(10^{-7})$        & $5.50^{+2.06}_{-1.69}$ &$5.8^{+2.4}_{-2.0}$ & $5.1\pm 0.5$  \\
$Br(B^-\to K^- \tau^+ \tau^-)(10^{-7})$  & $1.29^{+0.46}_{-0.37}$ &&   \\
$Br(B^-\to K^-  \nu  \bar{\nu})(10^{-6})$& $4.42^{+1.66}_{-1.35}$ &$4.4^{+1.5}_{-1.5}$  & $< 13$  \\ \hline\hline
$Br(\bar{B}_s^0\to K^+ l^- \bar{\nu}_l)(10^{-4})$       & $1.27^{+0.60}_{-0.43}$ & & $$  \\
$Br(\bar{B}_s^0\to K^+ \tau^-\bar{\nu}_\tau)(10^{-4})$  & $0.78^{+0.34}_{-0.27}$ & &  \\
$Br(\bar{B}_s^0\to K^0 l^+ l^-)(10^{-8})$             & $1.63^{+0.73}_{-0.58}$ &$1.4$ &  \\
$Br(\bar{B}_s^0\to K^0 \tau^+ \tau^-)(10^{-8})$       & $0.43^{+0.18}_{-0.15}$ &$0.3$&   \\
$Br(\bar{B}_s^0\to K^0 \nu  \bar{\nu})(10^{-7})$      & $1.31^{+0.58}_{-0.47}$ &$1.0$&   \\ \hline
\hline
 \end{tabular}
 \end{center}
\end{table}

In Table III, we list the NLO pQCD predictions for the branching ratios of
the $B_{(s)} \to (\eta,\etar)(l^+l^-,l\nu,\nu\bar{\nu})$ decays with $l=(e,\mu)$.
We considered two mixing schemes: (a) the traditional
Feldmann-Kroll-Stech (FKS) $\eta$-$\etar$ mixing scheme \cite{fks98,fks98b}
in the quark-flavor basis; and (b) the $\eta$-$\etar$-$G$ mixing scheme as defined
in Ref.~\cite{li-79a}: the physical states $\eta$, $\etar$
and $G$ are related to $\eta_q$, $\eta_s$, and $\eta_g$ through the rotation
matrix $U(\theta,\phi,\phi_G)$, which has been defined in Eq.~(4) of
Ref.~\cite{li-79a} with $\phi=\theta+54.7^\circ$ and $\phi_G\sim 30^\circ$.

In Table \ref{tab-brn2}, furthermore,  we show the pQCD predictions in the FKS mixing scheme
in column two  ( the NLO predictions with the total errors).
In Table III we also show the NLO pQCD predictions for the
branching ratios in the $\eta$-$\etar$-$G$ mixing scheme with the choice
of the mixing angles: $(\phi,\phi_G)=(43.7^\circ,33^\circ)$, the same
as in Ref.~\cite{liu-86R}. Currently available two measured values \cite{pdg2012} are
$Br(B^-\to\eta l^- \bar\nu_l)=(0.39\pm 0.08)\times 10^{-4}$ and
$Br(B^-\to\etar l^- \bar\nu_l)=(0.23\pm 0.08)\times 10^{-4}$.

Table \ref{tab-brn2} also includes a comparison between our pQCD predictions
and other theoretical results \cite{kim2001,chen07,chen10,choi10,wu06,azizi10}
obtained by using the different theoretical methods or approaches, including
for example the LCSR \cite{pball-98,pball-98b,kr00}, light-front quark model (LFQM)
\cite{choi10}, the lattice QCD\cite{Le2011,BL2013}.
One can see from the numerical results in Table \ref{tab-brn2} that the pQCD predictions
agree well with the theoretical predictions obtained from other nonperturbative methods.

%%%%%%%%%%%%%%%%%%%%%%%%%%%%%%%%%%%%%%%%%%%%%%%%vvv
\begin{table}[thb]
\begin{center}
\caption{The pQCD predictions at the NLO level in both the FKS $\eta$-$\etar$ mixing scheme
and the $\eta$-$\etar$-$G$ mixing scheme.  Other theoretical predictions
\cite{kim2001,chen07,chen10,choi10,wu06,azizi10} are listed in last column. }
\label{tab-brn2}
{\small
\begin{tabular}{l|cl|l}  \hline\hline
  Decay modes  &  $\eta$-$\etar$& $\eta$-$\etar$-$G$ & Others\\
\hline %%%%%%%%%%%%%
$Br(B^-\to\eta l^- \bar\nu_l)(10^{-4})$               & $ 0.41^{+0.15}_{-0.12}$ & $0.33^{+0.12}_{-0.10} $   &$0.43^{+0.08}_{-0.08}$[106]   \\
$Br(B^-\to\eta \tau^- \bar\nu_\tau)(10^{-4})$         & $ 0.24^{+0.09}_{-0.07}$ & $0.20^{+0.07}_{-0.05} $   & $0.29^{+0.07}_{-0.06}$[108] \\
$Br(B^-\to\eta^\prime l^- \bar\nu_l)(10^{-4})$        & $ 0.20^{+0.08}_{-0.06}$ & $0.16^{+0.06}_{-0.05} $   & $0.21^{+0.04}_{-0.04}$[106]\\
$Br(B^-\to\eta^\prime \tau^- \bar\nu_\tau)(10^{-4})$  & $ 0.10^{+0.04}_{-0.03}$ & $0.08^{+0.03}_{-0.02} $   &$0.13^{+0.03}_{-0.02}$[108]\\
%%%%%%%%%%%%%
$Br(\bar{B}^0\to\eta l^+ l^-)(10^{-8})$               & $0.48^{+0.16}_{-0.14} $ & $0.39^{+0.14}_{-0.11} $   & $0.6$[107]\\
$Br(\bar{B}^0\to\eta \tau^+ \tau^-)(10^{-9})$         & $0.98 ^{+0.33}_{-0.28}$ & $0.80^{+0.28}_{-0.22} $   &$1.1\pm 0.1$[109]\\
$Br(\bar{B}^0\to\eta \nu\bar\nu)(10^{-9})$            & $0.38^{+0.14}_{-0.12} $ & $0.31^{+0.11}_{-0.09} $   & \\
%%%%%%%%%%%%%
$Br(\bar{B}^0\to\eta^\prime l^+ l^-)(10^{-8})$        & $0.24^{+0.09}_{-0.07} $ & $0.19^{+0.07}_{-0.05} $   &$0.3$[107] \\
$Br(\bar{B}^0\to\eta^\prime \tau^+ \tau^-)(10^{-9})$  & $0.25^{+0.09}_{-0.07} $ & $0.20^{+0.07}_{-0.05} $   &  \\
$Br(\bar{B}^0\to\eta^\prime \nu\bar\nu)(10^{-9})$     & $0.18 ^{+0.07}_{-0.05}$ & $0.14^{+0.05}_{-0.04} $   &  \\
%%%%%%%%%%%%%
$Br(\bar{B}_s^0\to\eta l^+ l^-)(10^{-7})$             & $2.07 ^{+0.87}_{-0.72}$ & $2.59^{+1.09}_{-0.90} $   &$2.4$[100] \\
$Br(\bar{B}_s^0\to\eta \tau^+ \tau^-)(10^{-7})$       & $0.45 ^{+0.20}_{-0.16}$ & $0.56^{+0.25}_{-0.21} $   &$0.34$[109]\\
$Br(\bar{B}_s^0\to\eta \nu\bar\nu)(10^{-6})$          & $1.62 ^{+0.71}_{-0.55}$ & $2.03^{+0.89}_{-0.69} $   &$1.4$[110] \\
%%%%%%%%%%%%%
$Br(\bar{B}_s^0\to\eta^\prime l^+ l^-)(10^{-7})$     & $ 2.18 ^{+0.96}_{-0.76}$& $1.45^{+0.64}_{-0.50} $   &$1.8$[100] \\
$Br(\bar{B}_s^0\to\eta^\prime \tau^+\tau^-)(10^{-7})$& $ 0.27 ^{+0.12}_{-0.10}$& $0.18^{+0.07}_{-0.06} $   &$0.28$[110] \\
$Br(\bar{B}_s^0\to\eta^\prime \nu\bar\nu)(10^{-6})$  & $ 1.71^{+0.75}_{-0.60} $& $1.14^{+0.47}_{-0.40} $   &$1.3$[110] \\
\hline \hline
\end{tabular} }
\end{center}
\end{table}
%%%%%%%%%%%%%%%%%%%%%%%%%%%%%%%%%%%%%%%%%%%%%%%%vvv

Based on the theoretical predictions for the branching ratios of all considered
semileptonic decays of $B$ and $B_s$ meson as shown in
Table \ref{tab-brn1} and \ref{tab-brn2},  and the phenomenological
analysis presented in Refs.~\cite{Xiao-SL1,Xiao-SL2},
we have the following observations:
\begin{enumerate}
\item[(1)]
For the relevant transition form factors $F_{0,+,\rmt}(q^2)$, the
NLO pQCD predictions for their values and the $q^2$-dependence are consistent with those
obtained from the LCSR or other theoretical methods
\cite{pball-98,pball-98b,kr00,pball-05,jhep04-014,prd83-094031}.
The pQCD predictions for the NLO twist-2 contribution to the form factors is $\sim 20\%$
of the total value.

\item[(2)]
For the considered decay modes $\bar{B}^0 \to \pi^+l^-\bar{\nu}_l$,
$B^- \to \pi^0 l^-\bar{\nu}_l$, $\bar{B}^0\to\bar{K}^0 l^+ l^-$, and $B^- \to K^- l^+ l^-$, the
NLO pQCD predictions for their decay rates agree very well with currently available experimental
measurements.

\item[(3)]
From the direct comparison between the pQCD predictions for the branching ratio $Br(\bar{B}^0 \to \pi^+l^-\bar{\nu}_l)$
and the measured value,  the value of $V_{ub}$ can be extracted directly:
$|V_{ub}|= \left ( 3.80^{+0.66}_{-0.60}(th.)\pm 0.13 \right ) \times10^{-3}$.

\item[(4)]
For the branching ratios $Br(B^-\to \eta^{(')} l^-\bar{\nu}_l)$, the NLO twist-2 contribution to the
transition form factors can provide $\sim 25\%$ enhancement to the LO pQCD results, which leads
to a better agreement of the pQCD predictions with the measured values.

\item[(5)]
Analgous to the ratio $R(D)$, we here also defined several ratios of the branching ratios $R_\nu, R_C$ and $R_{N1,N2,N3}$,
calculated and listed the pQCD predictions for their values and the errors, these theoretical predictions will
be tested by the LHCb experiments and by the Super-B experiments in the near future.

\item[(6)]
For $B/B_s \to (\eta,\etar)(l^+l^-,l^-\bar{\nu}_l,\nu\bar{\nu})$ with
$l=(e,\mu,\tau)$ decays, we considered both the traditional FKS $\eta$-$\etar$
mixing scheme and the new  $\eta$-$\etar$-G mixing scheme, and we found
that the relevant pQCD predictions in these two mixing schemes are consistent with each other.

\end{enumerate}
%---------------------------------------------------------------------------------------------------------------%

\subsection*{3.3 $V_{ub}$ and $V_{cb}$, $B \to K^* \mu^+\mu^-$ decays }\label{sec:c3-3}

As is well-known, the best method to determine $|V_{ub}|$ ($|V_{cb}|$) is to measure
semileptonic decay rates for $B\to X_u l \nu$ ($B \to X_c l \nu$),
which is proportional to $|V_{ub}|^2$ ( $|V_{cb}|^2$ ).
By using the data from the inclusive or exclusive semileptonic decay modes, one can extract out those
two CKM elements simultaneously.
Since the experimental and theoretical techniques for
these inclusive and exclusive method are rather different and largely independent,
one can make a cross-check for the consistency of our understanding of the theory and the
experimental measurements.

For the experimental measurements of $V_{cb}$, it is now in good shape: the
values determined by the exclusive and inclusive processes become consistent.
In Ref.~\cite{fu2013}, for instance, Fu et al. calculated $B \to D$ transition
form factors by employing the QCD light-cone sum rule and then estimated the
value and the uncertainty of $|V_{cb}|$ from the data for the
semi-leptonic $B \to D l \bar{\nu}_l$ decays.
Their estimation for $|V_{cb}|$ shows a good agreement with the BABAR, CLEO and Belle
measurements.
For $V_{ub}$, however, it is still a puzzle, the tension between
the exclusive and inclusive values persists  at present.

%%----------------  B \to K^* \mu^+\mu^-

The $B^0\to K^{*0} \mu^+\mu^-$ decay is a self-tagging process with $K^{*0} \to K^+\pi^-$,
mediated by electroweak box and penguin type diagrams in the SM. The shape of the
angular distribution of the $(K^+\pi^-)\mu^+\mu^-$ system offers
particular sensitivity to the values of $C_{7\gamma}$ and $C_9$, and
to the contributions from the new physics beyond the standard model.
The differential decay rates of the considered decays also provides useful information on the
estimation about the contribution from those new particles appeared in the loops
but typically suffers from large theoretical errors
due to the large uncertainty of the hadronic form factors.
For the semileptonic decays $B \to K^{(*)} l^+ l^-$, furthermore,
there also exist non-factorizable QCD effects which can not be accounted for in form
factors or short-distance Wilson coefficients, both at small and large $Q^2$ region, as
discussed in Refs.~\cite{beneke01,beneke05,beylich11}.

In Ref.~\cite{lhcb1308a}, very recently, LHCb collaboration  reported their measurements for
the  differential branching fraction, $dB/dq^2$ of the $B^0 \to K^{*0} \mu^+\mu^-$ decay.
Measurements of the angular observables, $A_{FB}$
($A_T^{Re}$), $F_L$, $S_3$ ($A^2_T$) and $A_9$ have also been performed in the same $q^2$ bins.
The LHCb results \cite{lhcb1308a} are the most precise measurements of $dB/dq^2$ and the angular
observables to date.
The measured CP  asymmetries in $B^0\to K^{*0} \mu^+\mu^-$  \cite{lhcb-201308,lhcb-prl110}, for example, is of the form
\beq
{\cal A}_{CP}(B^0\to K^{*0} \mu^+\mu^-)=-0.072 \pm 0.041,
\eeq
which is consistent with the SM at $1.8\sigma$ \cite{lhcb-prl110}.

All of the observables are consistent with SM expectations and together put stringent
constraints on the contributions from new particles to $b \to s \mu^+\mu^-$ FCNC processes.
A bin-by-bin comparison of the measured angular distribution with the SM hypothesis indicates
an excellent agreement with p-values between $18\%$ and $72\%$.
The first LHCb measurement for the position of the zero-crossing point of the forward-backward (FB) asymmetry
for the decay mode $B^0\to K^{*0} \mu^+\mu^-$,
$q_0^2 = (4.9\pm 0.9)GeV^2/c^4$, agrees well with the SM prediction \cite{Bobeth12,Bobeth11,Ali06}:
$q_{0,SM}^2 \in  [3.9,4.4] GeV^2/c^4$.

For $B^0 \to K^{*0} \mu^+\mu^-$ decays, the previous measurements for the considered observables do
suffer from large theoretical errors due to the sizable uncertainties of relevant hadronic form factors.
The new observables $P_{4,5,6,7}$ as defined in Ref.~\cite{jhep1305-137}:
$P_i=S_i/\sqrt{F_L(1-F_L)}$, which have small form-factor uncertainties, especially at low
$q^2$ region. By using the full 2011 data sample, LHCb presented their first measurements
for these new observables \cite{lhcb1308b}.
For more details about theoretical studies
and experimental measurements of $B^0 \to K^{(*)} l^+l^-$ decays, one can see
Refs.~\cite{lhcb1308a,lhcb1308b,lhcb-prl110,Bobeth12,Bobeth11,Ali06,jhep1302-010} and references therein.

%%==================================================================================

\section*{4 \hspace{0.3cm} $B/B_s \to (D^{(*)},D_s^{(*)}) l^-\bar{\nu}_l$ decays}
\label{sec:c4}

In this section, we will present the pQCD predictions for the branching ratios
of $B/B_s \to (D^{(*)},D_s^{(*)}) l^-\bar{\nu}_l$ decays \cite{Xiao-SL3,Xiao-SL4},
and make some comparisons with those from the HQET method or other different approaches
~\cite{prl109-071802,prd85-114502,prl109-161801,
prd86-054014,jhep2013-01054,prd86-034027,prd86-114037,mpla27-1250183,Fajfer-1301}.

For $B \to D l\bar{\nu}_{\rm l}$ decays, the formulae of the differential decay rate
$d\Gamma(B \to D l\bar{\nu}_{\rm l})/dq^{\rm 2}$ can be obtained from
Eq.~(\ref{eq:dg01}) by simple replacements: $m_P\to m_D$ and $V_{ub} \to V_{cb}$.
For $B \to D^* l\bar{\nu}_{\rm l}$ decays,  the expressions of the
differential decay widths can be found in Refs.~\cite{lu-79,Xiao-SL3}.
For $B_s \to D_s^{(*)} l\bar{\nu}_{\rm l}$ decays, the formulae of the
differential decay rates can be found in Ref.\cite{Xiao-SL4}.

By using the relevant form factors as defined in Sec.2,
one can calculate directly the branching ratios for the considered decays
by the integrations over the whole range of $q^{\rm 2}$.
In Table \ref{tab:br38}, the pQCD predictions for the branching ratios of
the eight considered decay modes are listed in the column two. For the case of light leptons
$l=(e,\mu)$, we show the averaged results. In column three, we show the HQET predictions
obtained by our direct calculations using the formulaes as given in
Refs.~\cite{prd85-094025,jhep2013-01054}, which agree perfectly with those as given
in Ref.~\cite{prd85-094025}.
The measured values from BaBar \cite{prl109-101802} are also listed in last column
as an comparison.

In Table \ref{tab:ratios8}, we list the pQCD predictions for the values of the six $R(X)$ ratios
in the second column. As a direct comparison, we also show the HQET predictions calculated
by ourselves or those as given in  Refs.~\cite{prd85-094025},
other SM predictions as  presented in
Refs.~\cite{prl109-071802,prd85-114502,mpla27-1250183,Kosnik-1301},
and the BaBar measured values \cite{prl109-101802}.
Since the most hadronic and SM parameter uncertainties
are greatly canceled in the ratios of the corresponding branching ratios,
the theoretical errors of the pQCD predictions for R(X)-ratios are reduced
significantly to about $5\%$, similar in size with those in the HQET.

\begin{table}[thb]
\begin{center}
\caption{ The theoretical predictions for  $Br(B\to D^{(*)} l^- \bar{\nu}_{\rm l})$.
The measured values ~\cite{prl109-101802,prd79-012002,prd77-032002} are also
listed in last column. } \label{tab:br38}
\vspace{0.2cm}
\begin{tabular}{l ll  l} \hline \hline
~~ Channels~~ & pQCD(\%) \hspace{0.1cm} & HQET(\%)\hspace{0.1cm} &  BaBar(\%)  \\ \hline
$Br(\bar{B}^0 \to D^+ \tau^- \bar{\nu}_{\rm \tau})$ & $0.87^{+0.34}_{-0.28}$ & $0.63 \pm 0.06$        & $1.01 \pm 0.22$\\
$Br(\bar{B}^0 \to D^+ l^- \bar{\nu}_{\rm l})$       & $2.03^{+0.92}_{-0.70}$ &$2.13 ^{+0.19}_{-0.18}$ & $2.15 \pm 0.08$\\
$Br(B^- \to D^0 \tau^- \bar{\nu}_{\rm \tau})$       & $0.95^{+0.37}_{-0.31}$ &$0.69 \pm 0.06$         & $0.99 \pm 0.23$\\
$Br(B^- \to D^0 l^- \bar{\nu}_{\rm l})$             & $2.19^{+0.99}_{-0.76}$ &$2.30 \pm 0.20$         & $2.34\pm 0.14$\\ \hline
$Br(\bar{B}^0 \to D^{*+}\tau^-\bar{\nu}_{\rm \tau})$ & $1.36^{+0.38}_{-0.37}$ &$1.25\pm 0.04$ &$1.74\pm 0.23$\\
$Br(\bar{B}^0\to D^{*+} l^- \bar{\nu}_{\rm l})$      & $4.52^{+1.44}_{-1.31}$ &$4.94\pm 0.15$ &$4.69\pm 0.34$\\
$Br(B^-\to D^{*0} \tau^- \bar{\nu}_{\rm \tau})$      & $1.47^{+0.43}_{-0.40}$ &$1.35 \pm 0.04$&$1.71\pm 0.21$\\
$Br(B^- \to D^{*0} l^- \bar{\nu}_{\rm l})$           & $4.87^{+1.60}_{-1.41}$ &$5.35 \pm 0.16$&$5.40\pm 0.22$\\
\hline \hline
\end{tabular}\end{center} \end{table}

\begin{table*}
\centering
\caption{ The theoretical predictions for the six R-ratios obtained by employing the
pQCD approach  or other theoretical methods, and the measured values \cite{prl109-101802}.}
\label{tab:ratios8}\vspace{0.2cm}
\begin{tabular}{l ll lllll} \hline \hline
Ratio &pQCD & HQET &HQET\cite{prd85-094025}& SM \cite{prl109-071802,prd85-114502}
&SM \cite{mpla27-1250183} & SM \cite{Kosnik-1301} & BaBar \cite{prl109-101802} \\ \hline
$R(D^0)$   &$0.433^{+0.017}_{-0.027}$&$0.297^{+0.017}_{-0.016}$&$-$&$-$&$-$&$-$&$0.429\pm 0.097$ \\
$R(D^+)$   &$0.428^{+0.023}_{-0.033}$&$0.297\pm 0.017$&$-$&$-$&$-$&$-$&$0.469\pm 0.099$ \\
$R(D^{*0})$&$0.302^{+0.012}_{-0.014}$&$0.253\pm 0.004$&$-$&$-$&$-$&$-$&$0.322\pm 0.039$ \\
$R(D^{*+})$&$0.301^{+0.012}_{-0.015}$&$0.252\pm 0.004$&$-$&$-$&$-$&$-$&$0.355\pm 0.044$ \\ \hline
${\cal R}(D)$  & $0.430^{+0.021}_{-0.026}$ & $0.297\pm 0.017$& $0.296^{+0.016}_{-0.016}$&$0.316$&$0.315$ & $0.31$&$0.440 \pm 0.072$\\
${\cal R}(D^*)$& $0.301^{+0.013}_{-0.013}         $ & $0.252\pm 0.004$& $0.252^{+0.003}_{-0.003}$&$ -$&$0.260$&$-$&$0.332 \pm 0.030 $\\
\hline\hline
\end{tabular}
\end{table*}

From the theoretical predictions as collected in Table \ref{tab:br38} and \ref{tab:ratios8}
we have the following observations:
\begin{itemize}
\item[(1)]
The pQCD predictions for ${\rm Br}(B \to D^{(*)}
l^-\bar{\nu}_{\rm l})$ agree well with other theoretical predictions based on different
methods and the measured values within one standard deviation.

\item[(2)]
The previous SM predictions for $R(D^{(*)})$ as given in
Refs.~\cite{prl109-071802,prd85-114502,mpla27-1250183,Kosnik-1301}
are consistent with each other within their errors,
but there still exist a clear discrepancy between
these predictions and the BaBar's measurements \cite{prl109-101802}.

\item[(3)]
For $R(D)$ and $R(D^*)$, the pQCD predictions agree very well with the data,
the BaBar's anomaly of $R(D^{(*)})$ are explained successfully
in the framework of the SM by using the pQCD factorization approach.

\end{itemize}

Analogous to $R(D^{(*)})$ ratios, we also defined new ratios
$R_{\rm D}^l$ and $R_{\rm D}^\tau$ \cite{Xiao-SL3} and found the pQCD predictions
\beq
R_{\rm D}^{\rm l} &\equiv& \frac{ {\cal B}(B \to D^+l^- \bar{\nu}_{\rm l} ) + {\cal B}(B \to D^0 l^- \bar{\nu}_{\rm l})}{ {\cal B}(B
\to D^{*+} l^- \bar{\nu}_{\rm l} )+ {\cal B}(B \to D^{*0} l^- \bar{\nu}_{\rm l})}\non
&=& 0.450^{+0.064}_{-0.051},\label{eq:rdt5a} \\
R_{\rm D}^{\rm \tau} &\equiv & \frac{ {\cal B}(B \to D^+ \tau^- \bar{\nu}_{\rm \tau} )+ {\cal B}(B \to D^0 \tau^- \bar{\nu}_{\rm \tau})}{
{\cal B}(B \to D^{*+} \tau^- \bar{\nu}_{\rm \tau} )+ {\cal B}(B \to D^{*0} \tau^- \bar{\nu}_{\rm \tau})}\non
&=&0.642^{+0.081}_{-0.070}.
\label{eq:rdt6a}
\eeq
These new ratios may be more sensitive to the QCD dynamics than the old ones
and therefore should be tested in the forthcoming experiments.

%%====================================================================================

Following the same procedure as for the cases of the decays $\bar{B}^0\to D^{(*)}l \bar{\nu}_l$,
one can estimate the decay rates of the four $\bar{B}^0_s\to D_s^{(*)}l \bar{\nu}_l$
decays, and the four $R(X)$ ratios of the corresponding branching ratios.
The pQCD predictions are listed in Table \ref{tab:br39} and \ref{tab:ratios9}.
As comparisons, Table \ref{tab:br39} also include the theoretical predictions
for the branching ratios from other SM methods: for example, the constituent
quark model \cite{epjc51-601},
the QCD sum rules \cite{prd78-054011}, the LCSRs or the covariant light-front
quark model (CLFQM) \cite{prd80-014005,prd82-094031} and other methods
\cite{jpg39-045002,prd87-034033}.
The errors of the pQCD predictions as given in Table \ref{tab:br39} and \ref{tab:ratios9}
are the combinations of the major theoretical errors come from the uncertainties of
$\omega_{B_s}=0.50\pm 0.05$ GeV ($\omega_{B}=0.40\pm 0.04$ GeV)
and $m_c=1.35\pm 0.03$ GeV, while those induced
by the variations of $f_{B_s}$ ($f_{B}$) and $|V_{cb}|$ are canceled
completely in the pQCD predictions for the $R(X)$-ratios.

\begin{table*}
\centering
\caption{ The pQCD predictions for the decay rates (in units of $10^{-2}$) of
the decay modes in consideration. The theoretical predictions are given in
Refs.~\cite{epjc51-601,prd78-054011,prd80-014005,prd82-094031,jpg39-045002,prd87-034033}
are listed as a comparison.}
\label{tab:br39}\vspace{0.2cm}
\begin{tabular}{l |c |c c  c c c} \hline \hline
 Channel & pQCD & CQM\cite{epjc51-601}& QCDSRs\cite{prd78-054011}
& \cite{prd80-014005,prd82-094031} & IAMF\cite{jpg39-045002} & RQM\cite{prd87-034033} \\
\hline %%
$Br(\bar{B}_{s}^0 \to D_{s}^+ \tau^- \bar{\nu}_\tau)$ &$0.84^{+0.38}_{-0.28}$ & $--$ & $--$ & $0.33^{+0.14}_{-0.11}$& $0.47-0.55$& $0.62 \pm 0.05$\\
$Br(\bar{B}_{s}^0 \to D_{s}^+ l^- \bar{\nu}_l)$ &$2.13^{+1.12}_{-0.77}$ &$2.73-3.00$ & $2.8-3.8$ & $1.0^{+0.4}_{-0.3}$& $1.4-1.7$& $2.1 \pm 0.2$\\
\hline %%
$Br(\bar{B}_{s}^0 \to D_{s}^{*+}\tau^-\bar{\nu}_\tau)$ &$1.44^{+0.51}_{-0.42}$ &$--$& $--$ & $1.3^{+0.2}_{-0.1}$& $1.2-1.3$& $1.3 \pm 0.1$\\
$Br(\bar{B}_{s}^0\to D_{s}^{*+} l^- \bar{\nu}_l)$ &$4.76^{+1.87}_{-1.49}$ &$7.49-7.66$& $1.89-6.61$ & $5.2 \pm 0.6$& $5.1-5.8$& $5.3 \pm 0.5$\\
\hline \hline
\end{tabular} \end{table*}

\begin{table}[thb]
\begin{center}
\caption{The pQCD predictions for the four $R(X)$ ratios for
$\bar{B}^0_s\to D_s^{(*)}l \bar{\nu}_l$ decays \cite{Xiao-SL4}.}
\label{tab:ratios9}
\vspace{0.2cm}
\begin{tabular}{ cc |c c} \hline \hline
$R(D_s)$& $R(D_s^*)$ &  $R_{D_s}^l$& $R_{D_s}^\tau $  \\ \hline
$0.392\pm 0.022$ & $0.302\pm 0.011$ & $0.448^{+0.058}_{-0.041}$ & $0.582^{+0.071}_{-0.045}$
\\ \hline \hline
\end{tabular}
\end{center} \end{table}

%%%%%%%%%%%%%%%%%%%%%%%%%%%%%%%%%%%%%%%%%%%%%%%%%%%%%%%%%%%%%

From the theoretical predictions as collected in Table \ref{tab:br38}-\ref{tab:ratios9}
one can find the following points:
\begin{enumerate}
\item[(1)]
For the pQCD predictions for all R(X) ratios of the branching ratios, due to the large
cancelation of the theoretical errors in the ratios, the total theoretical errors now become
less than $13\%$, much smaller than those for the branching ratios themselves.
All these ratios could be measured at the LHCb experiments or the Super-B experiments in the near future.

\item[(2)]
The ratio $R(D_{s})$ and $R(D_{s}^{*})$ are defined \cite{Xiao-SL4}
in the same way as the ratios $R(D^{(*)})$ in Refs.~\cite{prl109-101802,prd85-094025}.
These ratios generally measure the mass effects of heavy $m_\tau$ against the
light $m_e$ or $m_\mu$.

\item[(3)]
The new ratios $R_D^{l,\tau}$ and $R_{D_s}^{l,\tau}$ will measure the
effects induced by the variations of the form factors for
$\bar{B}^0 \to (D,D^*) $  and $\bar{B}_s^0 \to (D_s,D_s^*)$ transitions.
In other words, the new ratios $R_D^{l,\tau}$ and $R_{D_{s}}^{l,\tau}$
may be more sensitive to the QCD dynamics which
controls the $B/B_s  \to (D^{(*)},D_s^{(*)} )$ transitions than the
old ratios ratios $R(D^{(*)})$ and $R(D_{s}^{(*)})$.

\item[(4)]
On the limit of the $SU(3)_F$ flavor symmetry, the four ratios defined for
$\bar{B}^0_s\to D_s^{(*)}l \bar{\nu}_l$ decays should be very similar with the
corresponding ones for $B\to D^{(*)} l \bar{\nu}_l$ decays.
The pQCD predictions as listed in Table \ref{tab:ratios8} $-$ \ref{tab:ratios9}
do support this expectation. The breaking of $SU(3)_F$ flavor symmetry
is less than $10\%$.

\item[(5)]
At present, only the ratio $R(D)$ and $R(D^*)$ have been measured  by
Belle and BaBar \cite{prl109-101802,babar2008,belle2007,belle2010}.
In order to check if the BaBar's anomaly do exist or not for
$\bar{B}^0_s\to D_s^{(*)}l \bar{\nu}_l$ decays, and to test the
$SU(3)_F$ flavor symmetry among $\bar{B}^0_s\to D_s^{(*)}l \bar{\nu}_l$ and
$B\to D^{(*)} l \bar{\nu}_l$ decays,  we strongly suggest LHCb and the forthcoming
Super-B experiments to measure these four new ratios $R(D_s), R(D_s^*)$,
$R_{D_{s}}^l$ and $R_{D_{s}}^\tau$.

\end{enumerate}

\section*{5 \hspace{0.3cm} Summary and expectations} \label{sec:c5}

The semileptonic decays of $B/B_s$ mesons represent a very rich physics. The three or four final state
particles are  rather special, since they allow for a wealth of angular observables, decay
rates  and asymmetries: sensitive to new physics, experimentally clean signatures and
theoretically well predicted. The B factory experiments, the LHCb, CMS and ATLAS, and the
forthcoming Super-B factories. A large number of events have been collected, and much much more are expected !

In this short review, we present the current status about the theoretical and experimental studies
for some important semileptonic decays of $B/B_s$ mesons. We firstly gave a brief introduction for the
experimental measurements for some phenomenologically interesting channels, such as the improved measurements for
$B/B_s \to P (l^+l^-, l^-\bar{\nu}_l, \nu \bar{\nu})$ mainly from LHCb experiments, the
BaBar's $R(D)$ and $R(D^*)$ anomaly, the $P_5^\prime$ deviation for $B^0 \to K^{*0} \mu^+ \mu^-$ decay.
We then made a careful discussion about the evaluations for the form factors relevant for the considered
semileptonic decays in the popular methods, such as the LCSRs, the heavy quark effective theory and the new pQCD
factorization approach. We listed the pQCD predictions for the form factors $F_{0,+,T}(q^2)$, $V(q^2)$ and
$A_{0,1,2}(q^2)$ for $B/B_s\to (\pi, K, D,D^*)$ transitions in Table I. We made numerical comparisons
and found that  the pQCD predictions at the low $q^2$ region agree well with those from other methods.

In Sec.III and IV, using the form factors obtained from the $k_{\rm T}$ factorization formulism,
we calculated and then presented the LO and/or NLO pQCD predictions for the decay rates of all considered SL decays
of the $B$ and $B_s$ mesons, for example, the charged current $B/B_s \to (\pi, K, \etap, D^{(*)}, D_s^{(*)})l\nu$ decays,
the neutral current $B/B_s\to (\pi, K, \etap )(l^+l^-,\nu\bar{\nu})$ processes. We also made careful
phenomenological analysis for these pQCD predictions, compared them with those from different methods and the measured
values from BaBar, Belle, LHCb and other collaborations. We found, in general, the following points:
\begin{enumerate}
\item[(1)]
For all the considered $B/B_s \to (\pi, K, \eta, \etar)$ $(l^+l^-, l^-\bar{\nu}_l, \nu \bar{\nu})$ decays, the
pQCD predictions at the NLO level for their decays rates agree well with the measured values or those
obtained by using other popular but different theoretical methods.

\item[(2)]
For the two ratios $R(D)$ and $R(D^*)$, our pQCD predictions do agree very well with the
measured values as reported by BaBar collaboration,
the so-called BaBar's anomaly about the ratios $R(D^{(*)})$ are therefore explained successfully
in the framework of the pQCD factorization approach.

\item[(3)]
Besides the ratios $R(D^{(*)})$ and $R(D_s^{(*)})$, we defined several new ratios
$R_D^{l,\tau}$ and $R_{D_s}^{l,\tau}$, which will measure the
effects induced by the variations of the form factors for
$\bar{B}^0 \to (D,D^*) $  and $\bar{B}_s^0 \to (D_s,D_s^*)$ transitions.
The new ratios $R_D^{l,\tau}$ and $R_{D_{s}}^{l,\tau}$
may be more sensitive to the QCD dynamics which
controls the $B/B_s  \to (D^{(*)},D_s^{(*)} )$ transitions than the
old  ratios $R(D^{(*)})$ and $R(D_{s}^{(*)})$.  we therefore
strongly suggest LHCb and the forthcoming Super-B experiments to measure these four new ratios $R(D_s), R(D_s^*)$,
$R_{D_{s}}^l$ and $R_{D_{s}}^\tau$.

\end{enumerate}

As is well-known, the heavy flavor b physics is a powerful tool to make a precision test for the standard model theory
and for the searches for the signal or evidence of the new physics effects beyond the standard model.
Precision measurements for these decays an probe at mass scales not attainable with direct measurements
at the high energy frontiers. LHC itself is a flavor factory, other environments/experiments also providing
crucial ¡§flavoured¡¨ data. At present, most experimental measurements are consistent with the SM
expectations, but some hints or tensions have been seen in a few observables.

At present, the LHCb is the most sensitive heavy flavor physics experiment \cite{G1312}.
So far we have used its first two years sensitivity to rule out some new physics models. But we know there must be new physics.
We do believe that the heavy flavor b physics has a bright future ahead,  many more exciting results and much high
precision are expected,  not only with results on existing data, but also from outstanding prospects with future
facilities such as the LHCb upgrade and Super-B factories.
We do believe that, again, the heavy flavor b physics has brilliant present and ambitious long-term prospect!

%%=====================================================================

\begin{acknowledgments}
The authors would like to thank Li Hsiang-nan, L\"u Cai-Dian, Wang Yu-Ming, Shen Yue-Long  and Liu Xin
for many valuable discussions.
This work is supported by the National Natural Science Foundation of
China under Grants No. 11235005 and 10975074.
\end{acknowledgments}

%---------------------------------------------------------------------------------------------------------------%

%========================= reference=========================%


\begin{thebibliography}{99}

\bibitem{Fajfer13a}
Fajfer S  (2013) New physics in $B\to D^{(*)} \tau \nu_\tau$ decay.
Talk given at Helmholtz International School ¡°Physics of Heavy Quarks¡±,July 15-28, 2013, JINR, Dubna, Russia


\bibitem{Ligeti13a}
Ligeti Z (2013) Flavour physics and CP violation. Talks presented at SSI 2013, July 8-19, SLAC, US

\bibitem{Gershon13a}
Gershon T (2013) Flavour Physics and CP Violation. Talks presented at CERN Summer Student Lecture Programme,
Aug.1-6, 2013, CERN, Geneva

\bibitem{Straub13a}
Straub D M (2013) Heavy Flavour Theory. Lectures given at CERN-FermiLab HCPSS, 2013, CERN, Geneva

\bibitem{prl109-101802}
Lees J P et al  BABAR Collaboration (2012)
Evidence for an Excess of $\bar{B} \to D^{(*)} \tau \bar{\nu}_{\rm \tau}$ Decays.
\prl 109: 101802

\bibitem{prd8807}  %% New
Lees J P et al  BABAR Collaboration (2013) Measurement of an excess of $\bar{B} \to D^{(*)} \tau \bar{\nu}_{\rm \tau}$
decays and implications for charged Higgs bosons. \prd 88:072012


\bibitem{prl109-071802}
Bailey J A  et al (2012) Refining new-physics searches in $B \to D\tau \nu$ decay with lattice QCD.
\prl  109(07): 071802

\bibitem{prd85-114502}
Bailey J A et al Fermilab Lattice and MILC Collaborations (2012)
$B_{\rm s}\to D_{\rm s}/B\to D$ semileptonic form-factor ratios and their application to
$BR(B^{\rm 0}_{\rm s} \to \mu^+\mu^-$.
\prd 85(11):114502; Errotem 86(03): 039904

\bibitem{prl109-161801}
Fajfer S, Kamenik J F, Nisandzic I et al (2012) Implications of Lepton Flavor Universality Violations in B Decays.
\prl 109(16):161801


\bibitem{prd86-054014}
Crivellin A, Greub C, Kokulu A (2012) Explaining $B\to D \tau \nu, B\to D^* \tau \nu$ and $B\to \tau \nu$ in a 2HDM of type III.
\prd 86(05):054014

\bibitem{jhep2013-01054}
Celis A, Jung M, Li X Q et al (2013)
Sensitivity to charged scalars in $B\to D \tau \nu$ and $B\to \tau \nu$ decays.
\jhep 2013(01): 054

%%-----------------------------------  11

\bibitem{prd86-034027}
Datta A,  Duraisamy M, Ghosh D (2012) Diagnosing new physics in $b\to c \tau \nu$ decays in the light of the recent BABAR result.
\prd 86(03): 034027

\bibitem{prd86-114037}
Choudhury D, Ghosh D K, Kundu A (2012) B decay anomalies in an effective theory. \prd 86(11):114037

\bibitem{mpla27-1250183}
Faustov R N, Galkin V O (2012)
Exclusive weak B decays involving $\tau$ lepton in the relativistic quark model. \mpla 27(31): 1250183.

\bibitem{Fajfer-1301}
Fajfer S and Nisandzic I (2013) Theory of $B \to \tau \nu$ and $B \to D^* \tau \nu$.
Conference: C12-09-28; arXiv:1301.1167 [hep-ph]

\bibitem{lhcb1308a}
Aaij A et al. LHCb collaboration, (2013) Differential branching fraction and angular analysis of the decay
$B^0\to K^{*0}\mu^+\mu^-$. \jhep 08: 131

\bibitem{lhcb1308b}
Aaij A et al. LHCb collaboration, (2013) Measurement of form-factor independent
observables in the
decay  $B^0\to K^{*0}\mu^+\mu^-$. \prl 111:191801. arXiv:1308.1707[hep-ex].

\bibitem{Serra13}
Serra N, (2013) Studies of electroweak penguin transitions of $b \to s \mu\mu$. LHCb-TALK-2013-208.
Talk given at EPSHEP 2013, 18-24 July, 2013, Stockholm, Sweden

\bibitem{alb2013}
Albrecht J (2013) Heavy Flavour Experiment. Lectures presented at CERN-FermiLab HCPSS-2013, Aug.28 - Sept.6,
2013, CERN, Geneva

\bibitem{Jager1305}
J\"ager S, Camalich J M (2013) On $B\to Vll$ at small dilepton invariant mass, power corrections, and new physics.
\jhep 05:043. arXiv:1212.2263[hep-ph].

\bibitem{gmvprd88}
Descotes-Genona S, Matiasb J, Virtob J (2013) Understanding the $B\to K^* \mu^+\mu^-$ Anomaly.
\prd 88:074002. arXiv:1307.5683[hep-ph].

%%-----------------------------------  21
\bibitem{as1308}
Altmannshofera W, Straub D M (2013) New Physics in $B\to K^* \mu^+\mu^-$.
\epjc 73: 2646

\bibitem{ggh1308}
Gaulda R, Goertzb F, Haischc U (2014) On minimal $Z'$ explanations of the
$B\to K^* \mu^+\mu^-$ anomaly. \prd 89: 015005

\bibitem{babar-83-032007}
del Amo Sanchez  P et al. BABAR Collaboration (2011)
Study of $B\to \pi l\bar{\nu}$ and $B \to \rho l\bar{\nu}$ decays and
determination of $|V_{ub}|$. \prd 83: 032007

\bibitem{babar11}  %% New
~del~Amo Sanchez P et al. BABAR Collaboration (2011)  Measurement of the
$B^0\to \pi^- l^+\nu$ and $B^+ \to \etap  l^+ \nu$ branching fractions,
the $B^0\to \pi^- l^+\nu$ and $B^+ \to \eta  l^+ \nu$ form-factor shapes,
and determination of $|V_{ub}|$. \prd 83:052011

\bibitem{cleo-99-041802}
Adam  N E et al. CLEO Collaboration (2007) Study of Exclusive Charmless Semileptonic B
Decays and $|V_{ub}|$. \prl 99:041802

\bibitem{belle-648-139}
Hokuue T et al. Belle Collaboration (2007) Measurements of branching fraction and
$q^2$ distributions for $B\to \pi l \nu$ and $B\to \rho l \nu$ decays with
$B \to D^{(*)} l \nu$ tagging. \plb 648:139

\bibitem{babar-86a}
Lees  J P et al. BABAR Collaboration (2012) Measurement of branching fractions and rate asymmetries in the rare
decays $B\to K^{(*)} l^+l^-$.  \prd 86, 032012 (2012);

\bibitem{babar-86b}
del Amo Sanchez P et al. BABAR Collaboration (2010)
Search for the rare decay $B \to K \nu \bar{\nu}$. \prd 82:112002

\bibitem{lhcb-2012a}  %% 5
Aaij R et al. LHCb Collaboration (2012) First observation of the decay
$B^+ \to \pi^+ \mu^+\mu^-$. \jhep 12:125

\bibitem{D-2012}  %%
Dissertori G (2012) Experimental summary. Talk given at the Moriond QCD,
La Thuile, Italy, March 10-17, 2012.

%%-----------------------------------  31

\bibitem{Buras-2012}
Buras A J (2012) Hunting animalcula with flavor in the LHC era.
Talk presented at the Cracow Epiphany Conference, Jan.9-11, 2012, Cracow, Poland.

\bibitem{babar2008}
Aubert B  et al. BABAR Collaboration (2008) Observation of semileptonic decays
$B \to D^* \tau^- \bar{\nu}_{\rm \tau}$ and Evidence for $B\to D \tau^-\bar{\nu}_{\rm \tau}$.
\prl   100(02): 021801

\bibitem{belle2007}
Matyja A  et al Belle Collaboration (2007)
Observation of $ B^0 \to D^{*-} \tau^+ \nu_{\rm \tau}$ Decay at Belle.
\prl  99(19): 191807

\bibitem{belle2010}
Bozek A  et al Belle Collaboration (2010)
Observation of $ B^+ \to \bar{D}^{*0} \tau^+ \nu_{\rm \tau}$ and Evidence
for $B^+ \to \bar{D}^{*0} \tau^+ \nu_{\rm \tau}$ at Belle.
\prd 82(07): 072005


\bibitem{prd85-094025}
Fajfer S, Kamenik J F, Nisandzic I (2012)
On the $B \to D^* \tau \bar{\nu}_{\rm \tau} $ Sensitivity to New Physics.
\prd 85(09): 094025

%%----------------------------------------------- 35
\bibitem{bozek-2013}
Bozek A (for Belle Collaboration) (2013)
The $B \to \tau \nu$ and $ B \to \bar{D}^{(*)} \tau^+ \bar{\nu}_{\rm \tau}$ measurements.
talk given at FPCP 2013, May 3-6, 2013, Buzios, Breizl

\bibitem{dg1301}
Descotes-Genon S et al. (2013) Implications from clean observables for the binned analysis of
$B\to K^* \mu^+\mu^-$ at large recoil.   \jhep 1301: 048

\bibitem{pball-98}
Ball P (1998)  $B \to \pi$ and $B \to K$ transitions from QCD Sum Rules on the
Light-Cone. \jhep 09:005

\bibitem{pball-98b}
Ball P, Zwicky R (2001) Improved analysis of $B\to \pi e \nu$ from QCD sum rules
on the light cone.  \jhep 10:019

\bibitem{kr00}
Khodjamirian A et al. (2000) Predictions on $B \to \pi \bar{l}\nu_l$, $D \to
\pi \bar{l}\nu_l$, and $D \to K \bar{l}\nu_l$ from QCD light-cone sum rules.
\prd 62:114002

\bibitem{huang01} %%  added in reversed version
Huang T, Li Z H and Wu X Y (2001) Improved approach to the heavy to light form-factors
in the light cone QCD sum rules. \prd 63: 094001

\bibitem{wang03} %%  added in reversed version
Wang Z G, Zhou M Z and Huang T (2003) $B \pi$ weak form-factor with chiral current
in the light cone sum rules. \prd 67: 094006

\bibitem{pball-05}
Ball P, Zwicky R (2005) New results on $B \to \pi, K, \eta$ decay form factors
from light-cone sum rules. \prd 71: 014015

%%----------------------------------------------- 41

\bibitem{jhep04-014}
Duplancic G, Khodjamirian A, Mannel T et al. (2008) Light-cone sum rules for
$B \to \pi$ form factors revisited. \jhep 04:014

\bibitem{zuo06} %%  added in reversed version
Zuo F, Li Z H and Huang T (2006) Form Factor for $B \to  D l \nu$ in Light-Cone Sum Rules
With Chiral Current Correlator. \plb 641: 177

\bibitem{wu08} %%  added in reversed version
Wu X G, Huang T and Fang Z Y (2008) $SU_f(3)$-symmetry breaking effects of the $B \to K$
transition form-factor in the QCD light-cone sum rules. \prd 77: 074001

\bibitem{wu09} %%  added in reversed version
Wu X G and Huang T (2009) Radiative Corrections on the $B \to  P$ Form Factors with Chiral
Current in the Light-Cone Sum Rules. \prd 79: 034013

\bibitem{huang09} %%  added in reversed version
Huang T, Li Z H and Zuo F (2009) Heavy-to-light transition form factors and their
relations in light-cone QCD sum rules. \epjc 60: 63

\bibitem{prd83-094031}
Khodjamirian A, Mannel T, Offen N, Wang Y M, $B\to \pi l \nu$ width and
$|V_{ub} |$  from QCD Light-Cone Sum Rules. \prd 83:094031

\bibitem{fu2013} %%  added in reversed version
Fu H B, Wu X G, Han H Y et al. (2014) $V_{cb}$ from the semileptonic decay
$B\to D l \bar{\nu}_l$ and the properties of the $D$ meson distribution amplitude.
\npb 884:172-192. arXiv:1309.5723 [hep-ph].

\bibitem{HQET1}
Falk A F, Neubert M (1993) Second order power corrections in the heavy quark effective theory. 1.
Formalism and meson form-factors. \prd 47:2965-2981

\bibitem{HQET2}
Falk A F, Neubert M (1993) Second order power corrections in the heavy quark effective theory. 2. Baryon
form-factors. \prd 47:2982-2990

\bibitem{HQET3}
Neubert M  (1994) Heavy quark symmetry. \pr 245:259-396

\bibitem{Grozin-2004}
Grozin A (2004) Heavy Quark Effective Theory, Springer, STMP 201

\bibitem{HPQCD-2006}
Gulez E et al. (2006)  B meson semileptonic form-factors from unquenched lattice QCD.
\prd  73: 074502

\bibitem{Le2011} 	
Lellouch L (2011) Flavor physics and lattice quantum chromodynamics.
Conference: C09-08-03.6, Proceedings p.629-698, arXiv:1104.5484

\bibitem{BL2013}
Bouchard C, Lepage G P et al. (HPQCD) (2013) $B_{(s)}$ semileptonic decays.
Talk at Lattice 2013, July 29- Aug. 3, 2013, Mainz, Germany

%%---------------------------------------------------  50
\bibitem{li-65}
Kurimoto T, Li H N, Sanda A I (2001) Leading power contribution to $B \to \pi, \rho$ transition form factors.
\prd 65: 014007(2001).

\bibitem{yang-npb642}
Wei Z T, Yang M Z (2002) The systematic study of $B \to \pi$ form factors in pQCD approach and its reliability.
\npb 642: 263

\bibitem{yang-epjc23-28}
L\"u C D, Yang M Z (2002) $B \to \pi \phi, \pi \omega$ decays in perturbative QCD approach.
\epjc 23: 275

\bibitem{lu-79}
Li R H, L\"u C D, Wang W et al. (2009) $B \to S$ transition form factors in the pQCD approach.
\prd 79: 014013

\bibitem{huang-71} %%  added in reversed version
Huang T, Wu X G (2005) Consistent analysis of the $B \to \pi$ transition form factor
in the whole physical region.  \prd 71: 034018

\bibitem{wu07} %%  added in reversed version
Wu X G, Huang T and Fang Z Y (2007) $B \to K$ transition form-factor up to $O(1/m(b)**2)$
within the $\kt$ factorization approach. \epjc 52:561

\bibitem{krw98}
Khodjamirian A, Ruckl R, Winhart C W (1998) Scalar $B \to \pi$ and $D \to \pi$
form factors in QCD. \prd 58:054013

\bibitem{npb592-3}
Beneke M, Feldmann Th (2001) Symmetry-breaking corrections to heavy-to-light B meson form
factors at large recoil. \npb 592:3-34

\bibitem{li-95}
Yeh T W, Li H N (1995) PQCD analysis of exclusive charmless B decay spectrum.
\plb 353:301

\bibitem{li-96}
Yeh T W, Li H N (1996) Perturbative QCD analysis of B meson decays. \prd 53:2480

\bibitem{li-97}
Yeh T W, Li H N (1997) Factorization theorems, effective field theory,
and nonleptonic heavy meson decays.
\prd 56, 1615 (1997);

\bibitem{Botts89}
Botts J, Sterman G (1989)  Hard Elastic Scattering in QCD: Leading Nehavior.
\npb 325:62
%%----------------------------------------------------------  60

\bibitem{Catani89}
Catani S, Trentadue L (1989) Resummation of the QCD Perturbative Series for
Hard Processes. \npb 327: 323

\bibitem{Sterman92}
Li~H N, Sterman~G (1992) The perturbative pion form factor with Sudakov suppression.
\npb 381: 129-140

\bibitem{Huang91}
Huang~T, Shen~Q~X (1991) The applicability of perturbative QCD to the pion form factor
and the pionic wavefunction. \zpc 50: 139-144

\bibitem{li-02}
Li H N, (2002) Threshold resummation for exclusive B meson decays.
\prd 66: 094010


\bibitem{Cao95}
Cao~F~G, Huang~T, Luo~C~W (1995) Reexamination of the perturbative pion
form factor with Sudakov suppression.
\prd 52: 5358

\bibitem{cheng06} %%  added in reversed version
Cheng H Y, Chua C K and Yang K C (2006) Charmless hadronic B decays involving scalar
mesons: Implications to the nature of light scalar mesons.
\prd 73: 014017

\bibitem{wal08}  %%  added in reversed version
Wang Y M, Aslam M J and Lu C D (2008) Scalar mesons in weak semileptonic decays
of $B_{(s)}$. \prd 78: 014006

\bibitem{han2013} %%  added in reversed version
Han H Y, Wu X G, Fu H B et al.  (2013)
Twist-3 Distribution Amplitudes of Scalar Mesons within the QCD Sum Rules and
Its Application to the $B \to S$ Transition Form Factors.
\epja 49: 78; arXiv:1301.3978 [hep-ph].


\bibitem{li-85074004}
Li H N, Shen Y L,  Wang Y M (2012) Next-to-Leading order corrections to $B \to \pi$
form factors in $k_{\rm T}$ factorization. \prd 85: 074004

\bibitem{cheng14a} %% added in reversed version
Cheng S, Yu X, Fan Y Y et al.  (2014) The NLO twist-3 contributions to
$B \to \pi$ form factors in $k_{T}$ factorization. In preparation.
\prd 89: 094004. arXiv: 1402.5501 [hep-ph]

\bibitem{Xiao-SL1}  %%  81
Wang W F, Xiao Z J (2012) Semileptonic decays $B/B_s \to (\pi, K)(l^+l^-,l\nu,\nu\bar{\nu} )$
in the perturbative QCD approach beyond the leading-order.
\prd 86: 114025

\bibitem{Xiao-SL2}
Wang W F, Fan Y Y, Liu M et al. (2013) Semileptonic decays $B/B_s \to (\eta,\etar, G)(l^+l^-,l\bar{\nu},\nu\bar{\nu} )$
in the perturbative QCD approach beyond the leading order.
\prd 87: 097501

\bibitem{Xiao-SL3}
Fan YY, Wang W F, Cheng S et al. (2014) Semileptonic decays $B \to D^{(*)} l\nu$ in
the perturbative QCD factorization approach. \csb 59:125-132

\bibitem{Xiao-SL4}
Fan YY, Wang W F, Xiao Z J (2014) Study of $\bar{B}_s^0 \to (D_s^+,D_s^{*+}) l^-\bar{\nu}_l$
decays in the pQCD factorization approach. \prd 89: 014030.

%%----------------------------------------------------------  84

\bibitem{Xiao-08a}
Xiao Z J, Zhang Z Q, Liu X et al. (2008) Branching ratios and CP asymmetries of $B\to K \etap$ decays in the
perturbative QCD approach. \prd 78: 114001

\bibitem{Xiao-13a}
Fan Y Y, Wang W F, Cheng S et al. (2013) Anatomy of $B\to K\etap$ decays in different mixing schemes and effects
of next-to-leading order contributions in the perturbative QCD approach.
\prd 87: 094003

\bibitem{Xiao-12a}
Xiao Z J, Wang W F, Fan Y Y (2012) Revisiting the pure annihilation decays $B_s\to \pi^+\pi^-$
and $B^0\to K^+K^-$: The data and the perturbative QCD predictions.
\prd 85:094003

\bibitem{keum2001}
Keum Y Y, Li H N, Sanda A I (2001) Fat penguins and imaginary penguins in perturbative QCD.
\plb 504:6-14


\bibitem{keum2001b}
Keum Y Y, Li H N, Sanda A I (2001) Penguin enhancement and $B\to K\pi$ decays in perturbative QCD.
\prd  63: 054008

\bibitem{lu2001}
C.D. L\"u, K. Ukai, and M.Z. Yang (2001) Branching ratio and CP violation of $B\to \pi\pi$ decays in the perturbative QCD approach.
\prd 63: 074009

\bibitem{xiao2006}
Liu X, Wang H S, Xiao Z J et al. (2006) Branching ratio and CP Asymmetry of $B\to \rho \etap$ decays in the
perturbative QCD approach.  \prd 73: 074002

\bibitem{pball-06}
Ball P, Braun V M, Lenz A (2006) Higher-twist distribution amplitudes of the K meson in QCD.
\jhep 05:004

\bibitem{pball-pi}
Ball P (1999)  Theoretical Update of Pseudo-S-Meson DAs of Higher Twist:
The Nonsinglet Case. \jhep 01:010

\bibitem{jhep1401-004} %% added in reversed version
Li H N, Shen Y L and Wang Y M (2014) Joint resummation for pion wave function and pion transition form factor.
\jhep 1401: 004

\bibitem{qiao01}  %% added in reversed version
Kawamura H, Kodaira J, Qiao C F et al. (2001) B meson light cone distribution
amplitudes in the heavy quark limit. \plb 523: 111

\bibitem{huang05} %% added in reversed version
Huang T, Wu X G and Zhou M Z (2005) B-meson wave function in the
Wandzura-Wilczek approximation. \plb 611: 260

\bibitem{huang06} %% added in reversed version
Huang T, Qiao C F and Wu X G (2006) B-meson wavefunction with 3-particle Fock states'
contributions. \prd 73: 074004

\bibitem{jhep1302-008} %% added in reversed version
Li H N, Shen Y L and Wang Y M (2013) Resummation of the rapidity logarithms in B meson wave functions.
\jhep 1302: 008

\bibitem{wh2013} %% added in reversed version
Wu X G and Huang T (2014) Heavy and light meson wave functions. 
\csb 59: 3801-3814. arXiv:1312.1455 [hep-ph].

%%----------------------------------------------------------  80

\bibitem{prd78-014018}
Li R H, L\"{u} C D, Zou H (2008) $B(B_s) \to  D_{(s)}P, D_{s}V, D^*_{(s)}P, D^*_{s}V$
decays in the pQCD approach. \prd 78(01): 014018

\bibitem{prd65-014007}
Kurimoto T, Li H N, Sanda A I (2001) Leading power contribution to $B \to \pi \rho$ transition form factors.
\prd 65(01): 014007

\bibitem{Bobeth-12}
Bobeth C (2012) Theory status of $b\to s l^+l^-$ decays and their combined analysis.
Talk given at Heavy Quarks and Leptons 2012, June 14, 2012, Prague. %% B\to K^* \mu^+\mu^-


\bibitem{Gallo-08}
Gallo F (2008) Study of exclusive charmless semileptonic decays of the
B meson in BaBar. ph.D Thesis, Universita' di Torino, 2008

\bibitem{BLT-12}
Bernlochner F U, Ligeti Z, Turczyk S (2012) A proposal to solve some puzzles in
semileptonic B decays. \prd 85:094033   %% B \to D, D^*


\bibitem{li1995}
Li H N (1995) Applicability of perturbative QCD to $B \to D$ decays.
\prd 52: 3958

\bibitem{prd67-054028}
Kurimoto T, Li HN, Sanda A I (2003) $B \to D^{(*)} $ form factors in perturbative QCD.
\prd 67: 054028

\bibitem{param-bz2}
Ball P (2007) $|V_{ub}|$ from UT angles and $B \to \pi l \nu$. \plb 644:38-44

\bibitem{jhep1009-089}
Khodjamirian A, Mannel T, Pivovarov A et al. (2010) Charm-loop effect in $B \to K^{(*)}l^+l^-$
and $ B \to K^* \gamma$. \jhep 1009: 089

\bibitem{CLN-97}
Caprini I, Lellouch L, Neubert M (1998) Dispersive bounds on the shape of
$\bar{B}\to D^{(*)} l \bar{\nu}$  form-factors.  \npb 530:153-181
%%----------------------------------------------------------  90

\bibitem{hfag2012}
Heavy Flavor Averaging Group, Amhis Y  et al (2012)
Averages of b-hadron, c-hadron, and tau-lepton properties as of early 2012,
arXiv:1207.1158v3 [hep-ex]

\bibitem{Buras96}
Buchalla G, Buras A J, Lautenbacher M E (1996) Weak decays beyond leading logarithms.
\rmp 68:1125-1244

\bibitem{Col-2006}
Colangelo P et al. (2006) Exclusive $B\to K^{(*)}l^+l^-$, $B\to K^{(*)}\nu\nu$ and
$B\to K^* \gamma$ transitions in a scenario with a single universal extra dimension.
\prd 73:115006

\bibitem{prd81}
Colangelo P, De Fazio F, Wang W, (2010) $B_s\to  f_0(980)$ form factors and $B_s$ decays
into $f_0(980)$. \prd  81: 074001

\bibitem{pdg2012}
Particle Data Group, Beringer J et al (2012) \prd 86:010001

\bibitem{Bartsch09}
Bartsch M et al. (2009) Precision flavour physics with $B\to K \nu \bar{\nu}$ and $B\to K l^+l^-$.
\jhep 11: 011

\bibitem{choi10}
Choi H M (2010) Exclusive rare $B_s \to (K, \eta,\etar )l^+l^-$ decays in the
light-front quark model. \jpg 37: 085005

\bibitem{wang08b}
Wang J J, Wang R M, Xu Y G et al. (2008) The Rare decays $B_u^+ \to \pi^+ l^+l^-, \rho^+l^+l^-$ and
$B_d^0 \to l^+l^-$ in the R-parity violating supersymmetry. \prd 77: 014017

\bibitem{lhcb-prd86a}
Lee J P et al. (2012) Branching fraction and form-factor shape measurements of exclusive charmless
semileptonic B decays, and determination of $|V_{ub}|$.
\prd 86: 092004

\bibitem{fks98}
Feldmann T, Kroll P, Stech B (1998) Mixing and decay constants of pseudoscalar mesons.
\prd 58:114006

\bibitem{fks98b}
Feldmann T, Kroll P, Stech B (1998) Mixing and decay constants of pseudoscalar
mesons: the sequel. \plb 449:339-346


\bibitem{li-79a}
Cheng H Y, Li H N, Liu K F (2009) Pseudoscalar glueball mass from $\eta-\eta'-G$ mixing.
\prd 79: 014024

%%----------------------------------------------------------  100


\bibitem{liu-86R}
Liu X, Li H N, Xiao Z J (2012) Implications on $\eta-\etar-$glueball mixing from
$B_{d/s}\to J/\psi\eta'$ Decays. \prd 86: 011501(R)

\bibitem{kim2001}
Kim C S, Yang Y D (2001) Study of the semileptonic decays
$B^\pm \to \eta^{(\prime)} l \nu$. \prd 65:017501


\bibitem{chen07}
Chen C H, Geng C Q (2007) $¦Ç^{(\prime)}$ productions in semileptonic B decays.
\plb 645:197-200


\bibitem{chen10}
Chen C H, Shen Y L, Wang W (2010) $|V_{ub}|$ and $B \to \etap$ form factors in
covariant light-front approach. \plb 686:118-123

\bibitem{wu06}
Wu Y L, Zhong M, Zuo Y B (2006) $ B_{(s)}, D_{(s)} \to \pi, K, \eta, \rho, K*, \omega,
\phi$  transition form factors and decay rates  with Extraction of the CKM parameters
$|V_{ub}|$, $|V_{cs}|$, $|V_{cd}|$. \ijmpa 21:6125-6172

\bibitem{azizi10}
Azizi K, Khosravi R, Falahati F (2010) Rare semileptonic $B_s$ decays to $\eta$ and $\etar$
mesons in QCD. \prd 82:116001

\bibitem{beneke01}
Beneke M, Feldmann T and Seidel D (2001) Systematic approach to exclusive
$B \to V l^+l^-, V\gamma$ decays. \npb 612: 25-58

\bibitem{beneke05}
Beneke M, Feldmann T and Seidel D (2005) Exclusive radiative and electroweak
$b \to d$ and $b \to s$ penguin decays at NLO.
\epjc 41: 173-188

\bibitem{beylich11}
Beylich M, Buchalla G and Feldmann T (2011) Theory of $B \to K^{(*)} l^+l^-$
decays at high $q^2$: OPE and quark-hadron duality.
\epjc 71: 1635

\bibitem{lhcb-201308}
Aaij R et al. LHCb Collaboration (2013) Differential branching fraction and angular analysis of
the decay $B^0\to K^{*0} \mu^+\mu^-$. \jhep 1308:131

\bibitem{lhcb-prl110}
Aaij R et al. LHCb collaboration (2013) Measurement of the CP asymmetry in $B^0\to K^{*0}\mu^+\mu^-$
decays. \prl 110: 031801

\bibitem{Bobeth12}
Bobeth C et al. (2012) The decay $\bar{B} \to \bar{K} l^+l^-$ at low hadronic recoil and model-independent $|\Delta B=1|$
constraints. \jhep 01: 107

\bibitem{Bobeth11}
Bobeth C, Hiller G, van Dyk D (2011) More benefits of semileptonic rare B decays at low recoil:
CP violation. \jhep 07: 067

\bibitem{Ali06}
Ali A, Kramer G, Zhu G H (2006) $ B\to K^* l^+l^-$ decay in soft-collinear effective theory.
\epjc 47: 625-641

\bibitem{jhep1305-137}
Descotes-Genon S, Hurth T, Matias J et al. (2013) Optimizing the basis
of $B \to K^* l^+l^-$ observables in the full kinematic range.
\jhep 1305:137

\bibitem{jhep1302-010}
Khodjamirian A, Mannel Th  and Wang Y M (2013) $B\to K l^+l^-$ decay at large hadronic recoil.
\jhep 1302:010

\bibitem{Kosnik-1301}
Becirevic D, Kosnik N, Tayduganov A (2012) $\bar{B} \to D\tau \bar{\nu}_{\rm \tau}$
vs. $\bar{B} \to D\mu \bar{\nu}_\mu$. \plb 716: 208-213

\bibitem{prd79-012002}
Aubert B  et al.  BABAR Collaboration (2009) Measurements of the semileptonic decays
$B\to D l\nu$ and $B \to D^* l\nu$ using a global fit to $DXl\nu$ final states.
\prd 79(01): 012002

\bibitem{prd77-032002}
Aubert B et al.  BABAR Collaboration (2008) Determination of the form factors for the decay
$B^0\to D^{*-} l\nu $ and of the CKM matrix element $V_{\rm cb}$.
\prd 77(03): 032002

\bibitem{epjc51-601}
Zhao S M, Liu X, Li S J (2007) Study of $B_s\to D_{sJ}(2317, 2460)l\bar{\nu}$
semileptonic decays in the CQM model.  \epjc 51:601-606

%%----------------------------------------------------------  110


\bibitem{prd78-054011}
Azizi K, Bayar M (2008)  Semileptonic $B_q \to D_q^* l\bar{\nu} (q = s, d, u)$
decays in QCD sum rules. \prd 78:054011

\bibitem{prd80-014005}
Li R H, Lu C D, Wang Y M (2009) Exclusive $B_s$ decays to the charmed mesons
$D_s^+(1968, 2317)$ in the standard model.  \prd 80: 014005

\bibitem{prd82-094031}
Li G, Shao F L, Wang W (2010) $B_s \to D_s(3040)$ form factors and $B_s$
decays into $D_s(3040)$. \prd 82: 094031


\bibitem{jpg39-045002}
Chen X J, Fu H F, Kim C S et al. (2012) Estimating form factors of $B_s \to D_s^{(*)}$
and their applications to semi-leptonic and non-leptonic decays.
\jpg 39: 045002

\bibitem{prd87-034033}
Faustov R N,  Galkin V O (2013) Weak decays of $B_s$ mesons to $D_s$ mesons in the
relativistic quark model. \prd 87: 034033

\bibitem{G1312}
G. Graziani (2013)  Highlights from LHCb. Talk given HEP 2013, Dec.16, 2013, Chile

%--------------------------------------------------------------------------------------------------
\end{thebibliography}
\end{document}